\documentclass[superscriptaddress,aps,letterpaper,10pt,twocolumn,floats,showpacs,amsmath,amsfonts,amssymb,pre,nofootinbib]{revtex4-2}
\usepackage{graphicx}
\usepackage[colorlinks=true,citecolor=blue]{hyperref}
\usepackage[caption=false]{subfig}
\usepackage{pstricks}
\usepackage{pst-node}
\usepackage{amsmath}
\usepackage{amsbsy}
\usepackage{verbatim}
\usepackage{dsfont}
\usepackage{MnSymbol}

\DeclareMathOperator*{\sgn}{sgn}

\begin{document}
	
	\title{Synchronization of thermodynamically consistent stochastic phase oscillators}

    \author{Maciej Chudak}
 \affiliation{Institute of Molecular Physics, Polish Academy of Sciences, Mariana Smoluchowskiego 17, 60-179 Pozna\'{n}, Poland}
	
	\author{Massimiliano Esposito}
	\affiliation{Complex Systems and Statistical Mechanics, Department of Physics and Materials Science, University of Luxembourg, 30 Avenue des Hauts-Fourneaux, L-4362 Esch-sur-Alzette, Luxembourg}

    \author{Krzysztof Ptaszy\'{n}ski}
 \email{krzysztof.ptaszynski@ifmpan.poznan.pl}
 \affiliation{Institute of Molecular Physics, Polish Academy of Sciences, Mariana Smoluchowskiego 17, 60-179 Pozna\'{n}, Poland}
	
	\date{\today}
	
	\begin{abstract}
We consider a toy model of two kinetically coupled stochastic oscillators whose dynamics is described as a Markov jump process among $N$ discrete phase states. For large $N$, it maps onto the deterministic two-oscillator Kuramoto model of synchronization. Despite its simplicity, we postulate its relevance for understanding more complex and realistic oscillator systems. In the thermodynamic limit, the model exhibits a continuous nonequilibrium phase transition between the unsynchronized and synchronized states. We show that this transition is not governed by any extremum dissipation principle -- depending on system parameters, synchronization may either reduce or enhance the dissipation. Close to the phase transition, we observe a divergent behavior of fluctuations and responses with $N$ and characterize their universal scaling behavior. In particular, the covariances of the oscillator phases and the local entropy productions are shown to diverge towards $-\infty$, a phenomenon that has not been reported before. Finally, we study the behavior of information-theoretic quantities, demonstrating that mutual information and information flow between oscillators display different scaling with $N$ in synchronized and unsynchronized states, and thus can act as order parameters of synchronization.
	\end{abstract}
	
	\maketitle

\section{Introduction}
Synchronization, a phenomenon in which coupled oscillators align their frequencies and phases, is widespread in many areas of science and technology~\cite{kuramoto1984chemical,pikovsky2001synchronization,strogatz2004sync}. Examples range from the synchronization of mechanical~\cite{pantaleone2002synchronization}, electronic~\cite{wiesenfeld1996synchronization} or chemical~\cite{marek1975synchronization} oscillators to the behavior of power grids~\cite{witthaut2022collective}, neurons~\cite{bick2020understanding} or the flashing of fireflies~\cite{strogatz2004sync}. The simplest model of synchronization is the Kuramoto model~\cite{kuramoto1975,kuramoto1984chemical,acebron2005kuramoto}, where the state of oscillators is described by a single phase variable, and their coupling is solely determined by the phase differences. Although such a description may appear simplistic, it is relevant for many types of weakly coupled oscillator systems (such as coupled metronomes~\cite{pantaleone2002synchronization} or arrays of Josephson junctions~\cite{wiesenfeld1996synchronization}), which can be mapped onto the Kuramoto model using phase reduction techniques~\cite{acebron2005kuramoto,nakao2016phase}. While the original Kuramoto model is deterministic, it has also been generalized to account for the influence of stochastic noise~\cite{acebron2005kuramoto,sakaguchi1988cooperative,majumder2025finite}.

From a physical perspective, coupled oscillators are typical examples of nonequilibrium systems that consume free energy and dissipate heat into the thermal environment in a way governed by the laws of thermodynamics. However, although the dynamics of synchronization has been widely studied in the literature, the associated thermodynamics has been comparatively less explored. In fact, the Kuramoto model is a purely dynamical model that does not provide any account of energy exchanges with the environment. Furthermore, noisy versions of the Kuramoto model usually just add noise ``by hand'' to deterministic equations~\cite{sakaguchi1988cooperative,acebron2005kuramoto}, which may lead to inconsistencies with the laws of thermodynamics~\cite{FalascoReview}.

The strategies for providing a thermodynamically consistent description of coupled oscillators are twofold. One strategy applies the Langevin equation by modeling oscillators as interacting particles driven by nonconservative forces~\cite{imparato2015stochastic,izumida2016energetics}. Here, we focus on another strategy, where the oscillators undergo a Markov jump process among discrete states of the system. Sometimes, such discretization is just a theoretical tool to add noise to a deterministic Kuramoto model~\cite{escaff2016synchronization,jorg2017stochastic}. However, in other cases, it is grounded in the microscopic physics of the model. The examples are chemical oscillators, whose microscopic dynamics corresponds to stochastic jumps between chemical configurations associated with chemical reactions~\cite{gaspard2002correlation,andrieux2008fluctuation,nandi2007effective,nandi2010intrinsic,zhang2020energy,zhang2025altruistic}. The thermodynamics of such discrete models can be described using the formalism of stochastic thermodynamics~\cite{Seifert2012,FalascoReview}, provided that all stochastic transitions are bidirectional. The thermodynamic description can be further grounded in microscopic models of their energetics and coupling to the thermal environment~\cite{herpich2018collective,herpich2019universality,meibohm2024minimum,meibohm2024small,zhang2020energy,zhang2025altruistic,gopal2025dissipation}.

The magnitude of noise in discrete-state oscillator systems is controlled by their size: they become effectively deterministic (noise-free) in an appropriate thermodynamic limit of an infinite system size. On the one hand, this limit can be achieved by increasing the number of coupled oscillators, leading to the emergence of deterministic limit cycles~\cite{
herpich2018collective,herpich2019universality,meibohm2024minimum,meibohm2024small,wood2006universality,wood2006critical,wood2007continuous,wood2007effects,assis2011infinite,assis2012collective,guislain2023nonequilibrium,gusilain2024discontinuous}. Here, we focus on the second strategy, where one increases the number of discrete states in a single oscillator, so that the dynamics of each of them becomes  deterministic~\cite{escaff2016synchronization,jorg2017stochastic,zhang2020energy,zhang2025altruistic}. These approaches can also be combined by taking both limits to obtain an analog of the infinite-oscillator Kuramoto model~\cite{jorg2017stochastic,zhang2020energy}.

The study of the thermodynamics of coupled oscillators has already provided some interesting results. For example, some coupled-oscillator models have been shown to obey maximum or minimum dissipation principles, so that, irrespective of their parameters, synchronization always enhances~\cite{zhang2020energy, gusilain2024discontinuous} or reduces~\cite{izumida2016energetics,
herpich2018collective,meibohm2024minimum,meibohm2024small,gopal2025dissipation} dissipation. 
Synchronization has also been shown to enhance thermodynamic performance in a model of coupled molecular motors~\cite{imparato2015stochastic}. Thermodynamics of synchronization has also been explored for quantum heat engines~\cite{jaseem2020quantum,murthado2023cooperation}, where it can enhance their power, efficiency~\cite{carrega2024dissipation}, or precision~\cite{razzoli2024synchronization}. However, many questions about thermodynamics and stochastic dynamics of coupled discrete-state oscillators remain unexplored.

In our study, we consider a toy model of two coupled stochastic oscillators whose dynamics is described as a Markov jump process among $N$ discrete phase states, with a preferred direction of jumps determined by a nonconservative thermodynamic force. The specific feature of our model is that the interaction between the oscillators affects only their kinetics, but not thermodynamic forces, in a way that is dependent only on the phase difference between the oscillators. In the thermodynamic limit $N \rightarrow \infty$, the model maps onto a two-oscillator Kuramoto model. Although we do not have any specific realization of such model in mind, we find it attractive for two reasons. First, its structure enables us to describe many features of its large-$N$ behavior analytically, which is rarely possible for more complex models. Second, it admits a very efficient numerical treatment, enabling the verification of our analytic results. Investigating our model, we particularly focus on the behavior at the transition between unsynchronized and synchronized states (shortly, the synchronization transition). We show that this is a genuine continuous nonequilibrium phase transition, associated with nonanalytic behavior of observables, and that this transition is not governed by any extremum dissipation principle. We also explore the finite-size scaling of responses and fluctuations, showing certain universalities. Finally, we analyze the behavior of information-theoretic quantities, mutual information and information flows, and show that they act as order parameters of synchronization (as suggested in Ref.~\cite{ameri2015mutual}). In our conclusions, we further justify why many of these observations should also be relevant for systems of coupled limit cycle oscillators, e.g., chemical oscillators~\cite{nandi2007effective,nandi2010intrinsic}.

The paper is organized as follows. In Sec.~\ref{sec:model} we describe the model considered. In Sec.~\ref{sec:phasedyn} we consider dynamics of average phases on both deterministic and stochastic level, including their finite-size scaling near the synchronization transition. In Sec.~\ref{sec:therm} we investigate the nonequilibrium thermodynamic behavior of the system. In Sec.~\ref{sec:fluct} we analyze the fluctuations of phases and entropy production. In Sec.~\ref{sec:inf} we investigate the behavior of information-theoretic quantities, mutual information and information flow. Finally, in Sec.~\ref{sec:concl} we present the conclusions that follow from our results.

\section{Model} \label{sec:model}

\begin{figure}[tbp]
    \centering
    \includegraphics[width=\linewidth]{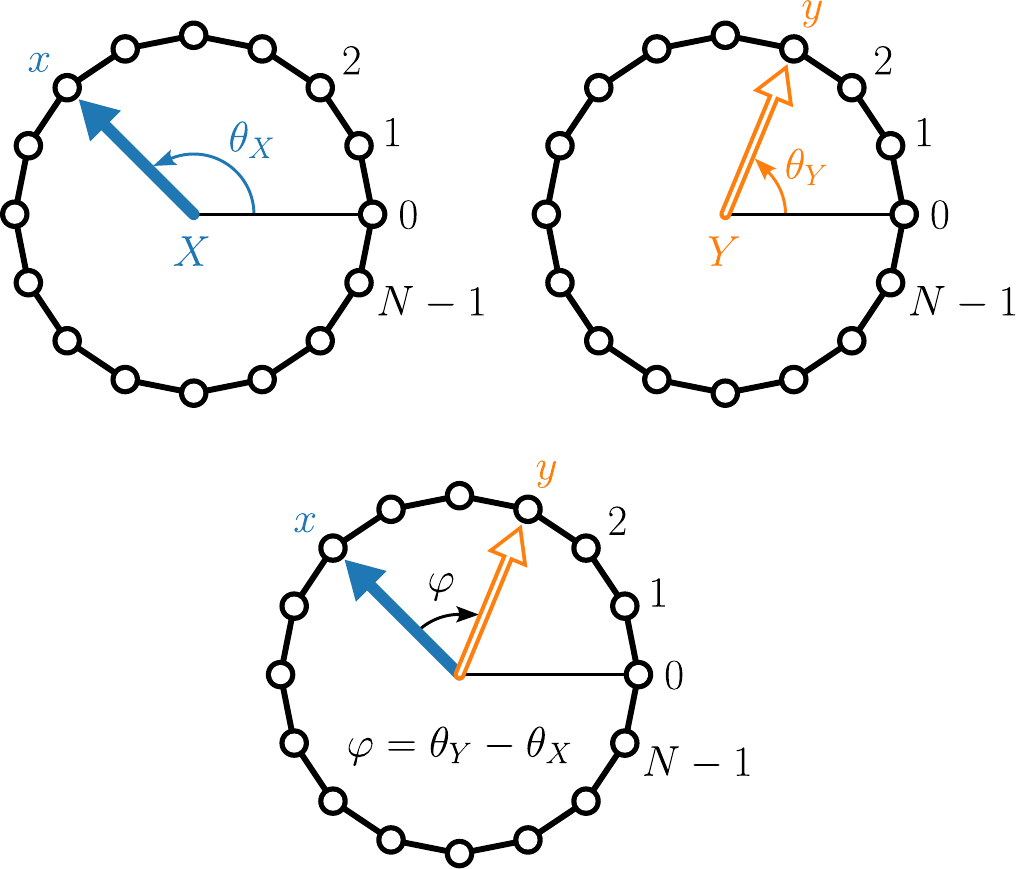}
    \caption{%
    Illustration of the model.
    Top: two coupled discrete-phase oscillators $X$ and $Y$, each consisting of $N$ discrete states, are in states $x$ and $y$. These states correspond to discrete phases $\theta_X=2 \pi x/N$ and $\theta_Y=2\pi y/N$. Bottom: the interaction between the oscillators depends only on the phase difference $\varphi \equiv \theta_Y-\theta_X = 2\pi(y-x)/N$.
    }
    \label{fig:model-illustration}
\end{figure}

We consider a model composed of two coupled discrete-phase oscillators labeled $X$ and $Y$ (see Fig.~\ref{fig:model-illustration}). Each oscillator is composed of $N$ discrete states, labeled $x$ and $y$, respectively, with $\{x,y \} \in \{0,\ldots,N-1\}$ defined modulo $N$. Then 
\begin{align} \label{eq:discphase}
\theta_X=2\pi x/N \,, \quad \theta_Y=2 \pi y/N \,,
\end{align}
are the discrete phases of the oscillators $X$ and $Y$, respectively. The joint probability of the state $(x,y)$ is denoted $p_{xy}$. It is further assumed that the system dynamics has a bipartite structure~\cite{horowitz2014thermodynamics}, that is, consists of stochastic jumps which change the states of either $X$ or $Y$ oscillators (i.e., there are no transitions that simultaneously change the states of both oscillators). The dynamics of the system is described by the master equation
\begin{align} \label{eq:masteq} \nonumber
 d_t p_{xy}=\sum_{\pm} & \left[ W^y_{x,x\pm 1} p_{x\pm 1,y}+W_{x}^{y,y \pm 1} p_{x,y \pm 1} \right. \\ & \left. -(W_{x \pm 1,x}^y+W_{x}^{y \pm 1,y}) p_{xy} \right] \,,
\end{align}
where 
$W^y_{x \pm 1,x}$ ($W_{x}^{y \pm 1,y}$) is the transition rate of a transition $x \rightarrow x \pm 1$ ($y \rightarrow y \pm 1$) for a fixed state $y$ ($x$). In our paper, we focus on the stationary state of the system where $d_t p_{xy}=0$.

To provide a well-defined thermodynamic description of the system, the transition rates are taken to obey the local detailed balance condition
\begin{align} \label{eq:locdetbal}
\ln \frac{W_{x \pm 1,x}^{y}}{W_{x,x \pm 1}^{y}}=\pm \beta f_X \,, \quad \ln \frac{W_{x}^{y \pm 1,y}}{W_{x}^{y,y \pm 1}}=\pm \beta f_Y \,,
\end{align}
where $\beta$ is the inverse temperature of the bath, while $f_X$ and $f_Y$ are nonconservative forces that enhance the transition rates in the direction of increasing phase. We further focus on a special parameterization of transition rates fulfilling this condition,
\begin{subequations} \label{eq:rate-parameterization}
 \begin{align}
 W_{x \pm 1,x}^y=N \Gamma_X \frac{1-a_X \sin \left[2\pi(x-y - \tfrac{1}{2} \pm \tfrac{1}{2})/N \right]}{1 + e^{\mp \beta f_X}} \,, \\
 W_{x}^{y \pm 1,y}=N \Gamma_Y \frac{1-a_Y \sin \left[2\pi(y-x - \tfrac{1}{2} \pm \tfrac{1}{2})/N \right]}{1 + e^{\mp \beta f_Y}} \,.
 \end{align}
\end{subequations}
Here, $\Gamma_X, \Gamma_Y$ are kinetic prefactors that parameterize the timescale of transitions, and $a_X,a_Y \in (-1,1)$ parameterize the effect of one subsystem on the dynamics of the other. We further take the rates to scale proportionally to $N$, so that the timescale of the oscillations remains finite for $N \rightarrow \infty$. For the toy model considered, this scaling is introduced artificially. However, its physical motivation comes from the fact that an extensive scaling of transition rates with system size is intrinsic to microscopic models of limit cycle oscillators~\cite{FalascoReview}, e.g., chemical oscillators~\cite{gaspard2002correlation,andrieux2008fluctuation} or systems of coupled oscillating units~\cite{
herpich2018collective,herpich2019universality,meibohm2024minimum,meibohm2024small,wood2006universality,wood2006critical,wood2007continuous,wood2007effects,assis2011infinite,assis2012collective,guislain2023nonequilibrium,gusilain2024discontinuous}.

Importantly, in contrast to previous works on the thermodynamics of synchronization, in the considered model, the coupling between subsystems does not affect the thermodynamic forces acting on a subsystem, i.e., it does not modify the right-hand side of the local detailed balance condition~\eqref{eq:locdetbal}. Instead, it affects only the system's kinetics, by modulating the symmetric parts of the transition rates $\tilde{W}^y_{x + 1,x}=\sqrt{{W}^y_{x + 1,x} {W}^y_{x,x + 1}}$ and $\tilde{W}_x^{y + 1,y}=\sqrt{{W}_x^{y + 1,y} {W}_x^{y,y + 1}}$. As later shown, this significantly simplifies the analysis of the system's thermodynamics. 

\subsection{Reduction to one-dimensional model} \label{subsec:reduction}
We now note that the transition rates \eqref{eq:rate-parameterization} depend only on the parameter $i=y-x \in \{0,\ldots,N-1\}$, defined modulo $N$, which plays the role of a discrete phase difference between two oscillators (see bottom of Fig.~\ref{fig:model-illustration}). As a result, we can reduce the original problem to an effective one-dimensional Markov jump process among the discrete phase difference states $i$, each occupied with a probability $p_i$. This is beneficial for both numerical calculations, enabling analysis of systems with very large $N$, and analytic calculations, allowing the use of methods suited for the description of one-dimensional models. The master equation corresponding to that reduced description reads
\begin{align} \label{eq:masteq-eff}
 d_t p_{i}=\sum_{\pm} & \left( W_{i,i\pm 1} p_{i \pm 1}-W_{i \pm 1,i} p_i \right) \,.
\end{align}
The transition rates can be decomposed as
\begin{align}
W_{i \pm 1,i}=W^X_{i \pm 1,i}+W^Y_{i \pm 1,i}  \,,
\end{align}
where $W^\alpha_{i \pm 1,i}$ denote the transition rates from state $i$ to $i\pm 1$ associated with transitions in the oscillator $\alpha$. Applying Eq.~\eqref{eq:rate-parameterization}, these transition rates can be expressed as
\begin{subequations} \label{eq:rate-parameterization-eff}
 \begin{align}
 W_{i \pm 1,i}^{X} &=N \Gamma_X \frac{1-a_X \sin \left[2\pi(-i - \tfrac{1}{2} \mp \tfrac{1}{2})/N \right]}{1 + e^{\pm \beta f_X}} \,, \\
 W^{Y}_{i \pm 1,i} &=N \Gamma_Y \frac{1-a_Y \sin \left[2\pi(i - \tfrac{1}{2} \pm \tfrac{1}{2})/N \right]}{1 + e^{\mp \beta f_Y}} \,.
 \end{align}
\end{subequations}
We then define the matrix $\mathbb{W}$ with off-diagonal elements $W_{ij}$ and diagonal elements $W_{ii}=-\sum_{j \neq i} W_{ji}$. The stationary state vector $\boldsymbol{p}=(\ldots,p_i,\ldots)^\intercal$ is then given by the stationary solution of the master equation
\begin{align}
d_t \boldsymbol{p}=\mathbb{W} \boldsymbol{p}=0 \,.
\end{align}
The stationary state probabilities in the original model can then be determined by noting that every state $(x,y)$ with a given $y-x=i$ is equally likely. Thus, $p_{xy}=p_i/N$.

\section{Deterministic and stochastic dynamics} \label{sec:phasedyn}

\subsection{Deterministic dynamics}
We now analyze the phase dynamics of the oscillators. We first consider the thermodynamic limit $N \rightarrow \infty$, where the behavior of the system becomes effectively deterministic. To that end, we consider the discrete phases defined via Eq.~\eqref{eq:discphase} to be continuous variables. In the thermodynamic limit, they can be shown to evolve according to deterministic mean-field equations~\cite{FalascoReview}
\begin{subequations}
\begin{align}
d_t \theta_X &=2 \pi\lim_{N \rightarrow \infty} \left(W_{x+1,x}^y-W_{x-1,x}^y \right)/N \,, \\
d_t \theta_Y &=2 \pi \lim_{N \rightarrow \infty} \left(W^{y+1,y}_x-W^{y-1,y}_x \right)/N \,.
\end{align}
\end{subequations}
Applying the parameterization~\eqref{eq:rate-parameterization}, we obtain
\begin{subequations} \label{eq:detphasedyn}
\begin{align}
d_t \theta_X &=\Omega_X-K_X \sin (\theta_X-\theta_Y) \, , \\
d_t \theta_Y &=\Omega_Y-K_Y \sin (\theta_Y-\theta_X) \,,
\end{align}
\end{subequations}
where
\begin{align} \label{eq:defomega}
 \Omega_\alpha \equiv 2 \pi \Gamma_\alpha \tanh(\beta f_\alpha/2) \,
\end{align}
is the intrinsic frequency of the oscillator $\alpha \in \{X,Y \}$, while 
\begin{align} \label{eq:defk}
 K_\alpha \equiv a_\alpha \Omega_\alpha\,
\end{align}
is the coupling parameter that characterizes the influence of the other oscillator on the dynamics of the oscillator $\alpha$. We note that Eq.~\eqref{eq:detphasedyn} corresponds to the two-oscillator version of the paradigmatic Kuramoto model of synchronization~\cite{kuramoto1975,acebron2005kuramoto,kuramoto1984chemical}. However, in our model, both $\Omega_\alpha$ and $K_\alpha$ depend on the nonconservative forces $f_\alpha$, so that they are nonzero only out of equilibrium (i.e., $f_\alpha>0)$. 
Furthermore, even when the parameters $a_\alpha$ and $\Gamma_\alpha$ are equal, the coupling is generically nonreciprocal, $K_X \neq K_Y$ (except when both forces are equal, $f_X=f_Y$, or much larger than the temperature, $\beta f_\alpha \gg 1$). We note that systems with such nonreciprocal couplings have recently attracted attention in multiple contexts~\cite{ivlev2015statistical,loos2020irreversibility,fruchart2021non,avni2025nonreciprocal,avni2025dynamical,meredith2020predator,reisenbauer2024non}, including synchronization~\cite{hong2011kuramoto,amro2015phase,aktay2024neuromodulatory,chakraborty2025effects,nadolny2025nonreciprocal,lai2025nonreciprocal}.

The dynamics of a single oscillator is characterized by its observed frequency $\bar{\Omega}_\alpha$, defined as the time-averaged phase velocity
\begin{align}
\bar{\Omega}_\alpha \equiv \lim_{\tau \rightarrow \infty} \frac{1}{\tau} \int_0^\tau dt d_t \theta_\alpha \,.
\end{align}
Due to the interaction between oscillators, the observed frequency can differ from the intrinsic frequency $\Omega_\alpha$. To determine this quantity, we introduce the parameters
\begin{align} \label{eq:parametersdet}
\varphi=\theta_Y-\theta_X \,, \;
\omega=\Omega_Y-\Omega_X \,, \;
K=K_X+K_Y \,,
\end{align}
that characterize the phase difference between the oscillators, the detuning of oscillator intrinsic frequencies, and the total coupling between the oscillators, respectively. Then, the dynamics of the phase difference $\varphi$ is given by the single-variable differential equation, called the \textit{Adler equation}~\cite{adler1946study},
\begin{align} \label{eq:adler}
d_t \varphi=F(\varphi) \equiv \omega-K \sin (\varphi) \,,
\end{align}
where we call $F(\varphi)$ a drift term.

\subsubsection{Synchronized state} 
The above equation has different solutions for $|\omega| \leq |K|$ and $|\omega| > |K|$. In the former case ($|\omega| \leq |K|$) the Adler equation has a stable fixed point
\begin{align} \label{eq:fixedpointadler}
\varphi^*=\arcsin(\omega/K) \,.
\end{align}
As a result, the oscillators are synchronized -- their phases evolve with the same constant velocity $d_t \theta_X=d_t \theta_Y$. Consequently, their observed frequencies $\bar{\Omega}_\alpha$ align to the common frequency 
$\bar{\Omega}$,
\begin{align}
\bar{\Omega}_X=\bar{\Omega}_Y=\bar{\Omega} \,,
\end{align}
which is given by the weighted average of intrinsic frequencies of both oscillators,
\begin{align} \label{eq:freqsync}
\bar{\Omega}=\frac{K_Y \Omega_X+K_X \Omega_Y}{K} \,.
\end{align}
This can be calculated by inserting Eq.~\eqref{eq:fixedpointadler} into Eq.~\eqref{eq:detphasedyn}.

\subsubsection{Unsynchronized state} 
For $|\omega| > |K|$, the Adler equation~\eqref{eq:adler} has a running periodic solution. Its period can be calculated as
\begin{align} \label{eq:periodphasedif}
T=\int_{0}^{2\pi} \frac{d \varphi}{F(\varphi)}=\frac{2 \pi \sgn(\omega)}{\sqrt{\omega^2-K^2}} \,.
\end{align}
Here, we employ the fact that $d\varphi/F(\varphi)=d \varphi/d_t \varphi$ is the amount of time after which the phase difference increases by $d \varphi$. We also apply the convention in which $T$ can be either positive or negative, depending on whether $\varphi$ increases or decreases. As a result, oscillators have different observed frequencies $\bar{\Omega}_\alpha$, which can be determined by averaging the phase velocity over a single period,
\begin{subequations}
\begin{align}
\bar{\Omega}_X &=\frac{1}{T} \int_0^T dt d_t \theta_X=\Omega_X+\frac{K_X}{T} \int_{0}^{2\pi} \frac{\sin(\varphi) d \varphi}{F(\varphi)} \,, \\
\bar{\Omega}_Y &=\frac{1}{T} \int_0^T dt d_t \theta_Y=\Omega_Y-\frac{K_Y}{T} \int_{0}^{2\pi} \frac{ \sin(\varphi) d \varphi}{F(\varphi)} \,,
\end{align}   
\end{subequations}
where in the second step we inserted Eq.~\eqref{eq:detphasedyn} and used $dt=d \varphi/d_t \varphi$. 
The solution yields
\begin{subequations} \label{eq:solomegaxdet}
\begin{align} 
\bar{\Omega}_X&=\bar{\Omega}-\frac{K_X}{K} \vartheta =\Omega_X+\frac{K_X}{K} (\omega-\vartheta)\,, \\
\bar{\Omega}_Y&=\bar{\Omega} +\frac{K_Y}{K}\vartheta =\Omega_Y-\frac{K_Y}{K} (\omega-\vartheta) \,,
\end{align}   
\end{subequations}
where
\begin{align} \label{eq:freqdetuning}
\vartheta \equiv \bar{\Omega}_Y-\bar{\Omega}_X =\frac{2\pi}{T}=\omega \sqrt{1-K^2/\omega^2} \,
\end{align}
is the detuning of the observed frequencies.

\subsubsection{No synchronization for opposite signs of $\Omega_X$ and $\Omega_Y$} 
We now note that in our model (in contrast to a generic Kuramoto model), the magnitude of the couplings $K_\alpha$ is bounded as $|K_\alpha| \leq |\Omega_\alpha|$ because $a_X,a_Y \in (-1,1)$ [see Eq.~\eqref{eq:defk}]. This puts certain constraints on the dynamics of the model. The first of them is that the oscillators cannot synchronize when the intrinsic frequencies $\Omega_X$ and $\Omega_Y$ are of opposite sign. This results from the inequality
\begin{align}
|K|=|a_X \Omega_X+a_Y  \Omega_Y| \leq |a_X \Omega_X|+|a_Y \Omega_Y| < |\Omega_X|+|\Omega_Y| \,,
\end{align}
where we used $a_X,a_Y \in (-1,1)$. For opposite signs of $\Omega_X$ and $\Omega_Y$, we have $|\Omega_X|+|\Omega_Y|=|\omega|$. Consequently, $|K| < |\omega|$, which precludes synchronizations.

\subsubsection{Maximum frequency shift -- no changes of direction of the oscillations} \label{subsec:nochanges}
The second universal constraint is that the maximum shift of the oscillator frequency is bounded as $|\bar{\Omega}_\alpha-\Omega_\alpha| \leq |K_\alpha|$. In our model, where $|K_\alpha|=|a_\alpha \Omega_\alpha| \leq |\Omega_\alpha|$, this implies that the interaction between oscillators does not change the direction of the oscillations, i.e., the observed frequency of the oscillations $\bar{\Omega}_\alpha$ has the same sign as the intrinsic frequency $\Omega_\alpha$. To prove that constraint, we note that for the synchronized state
\begin{align}
|\bar{\Omega}_\alpha-\Omega_\alpha| = |\bar{\Omega}-\Omega_\alpha|=|K_\alpha \omega/K| \leq |K_\alpha| \,,
\end{align}
where in the second step we used Eq.~\eqref{eq:freqsync} and $\Omega_\alpha=(K_X+K_Y)\Omega_\alpha/K$, and in the last step we used $|\omega/K| \leq 1$ in the synchronized state. In the unsynchronized state, from Eq.~\eqref{eq:solomegaxdet}, we have
\begin{align}
|\bar{\Omega}_\alpha-\Omega_\alpha|=\left|K_\alpha (\omega/K)\left(1-\sqrt{1-K^2/\omega^2} \right) \right| \leq |K_\alpha| \,,
\end{align}
where we used $|(\omega/K)(1-\sqrt{1-K^2/\omega^2} )| \leq 1$ for $|K/\omega| \leq 1$. 

\subsection{Comparison with the stochastic description}
We now compare the results of the deterministic approach with the numerical results for finite $N$. In the latter context, the observed frequency is a stochastic quantity defined as
\begin{subequations} \label{eq:stochfreq}
\begin{align}
\bar{\Omega}_X &\equiv \frac{2\pi}{N} \sum_{\pm} \sum_{x,y} \pm W^y_{x \pm 1,x} p_{xy} \,, \\ \bar{\Omega}_Y &\equiv \frac{2\pi}{N} \sum_{\pm} \sum_{x,y} \pm W^{y \pm 1,y}_x p_{xy} \,.
\end{align}
\end{subequations}
This expression comes from the fact that every jump $x \rightarrow x \pm 1$ ($y \rightarrow y \pm 1$) changes the phase $\theta_X$ ($\theta_Y$) by $\pm 2 \pi/N$ [see Eq.~\eqref{eq:discphase}]. Consequently, the observed frequency $\bar{\Omega}_X$ ($\bar{\Omega}_Y$) is proportional to the rate of jumps $x \rightarrow x+1$ ($y \rightarrow y+1$), minus the rate of opposite jumps, multiplied by the proportionality factor $2 \pi/N$. 

\begin{figure}
    \centering
    \includegraphics[width=0.95\linewidth]{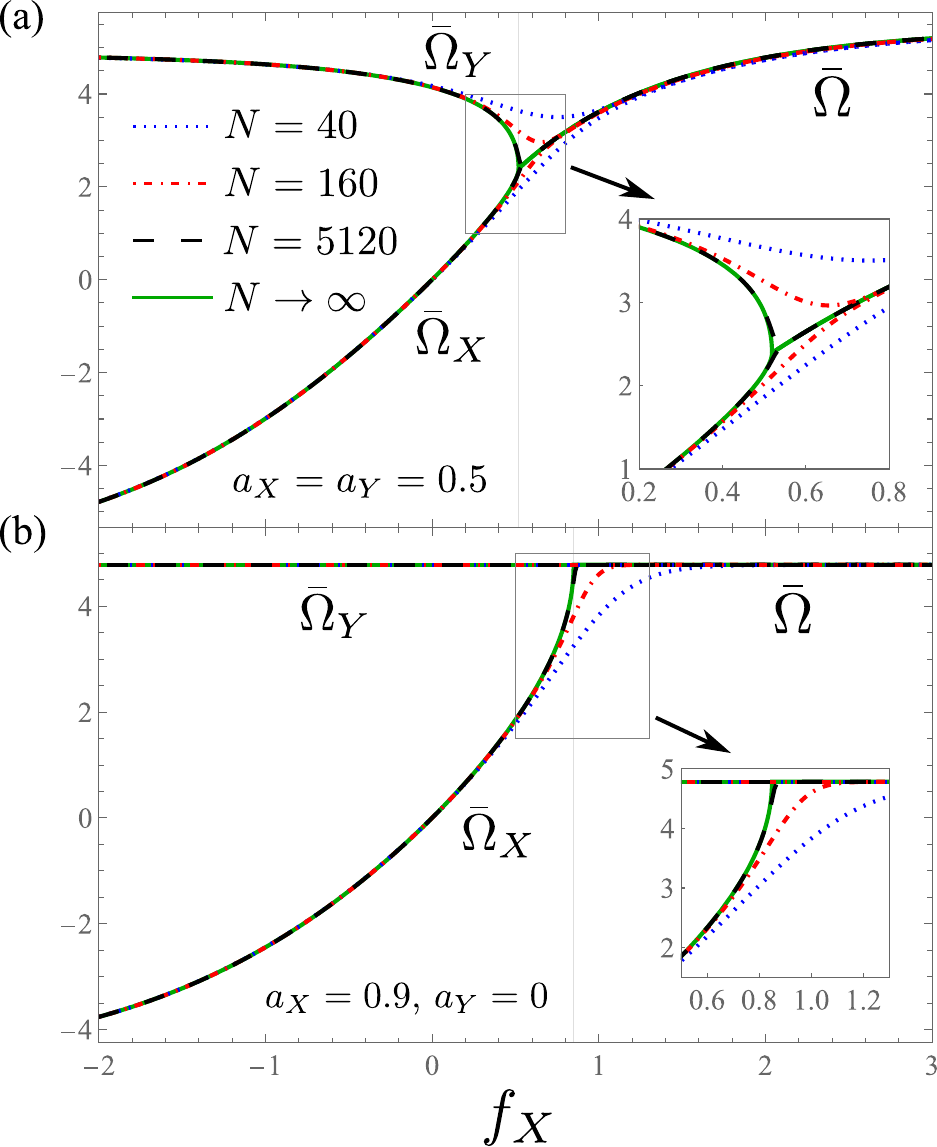}
    \caption{The observed frequencies $\bar{\Omega}_\alpha$ for (a) the symmetric coupling $a_X=a_Y=0.5$ and (b) the asymmetric coupling $a_X=0.9$, $a_Y=0$. The finite-size results are given by the master equation, while the results for $N \rightarrow \infty$ are given by the deterministic approach. The vertical gray lines denote the phase transition from unsynchronized (left) to synchronized (right) state. The insets show the behavior of $\bar{\Omega}_\alpha$ close to the phase transition point, in the region enclosed by a small rectangle. Other parameters:  $f_Y=2$, $\Gamma_X=\Gamma_Y=1$, $\beta=1$.}
    \label{fig:freq}
\end{figure}

The comparison of the outcomes of the deterministic and stochastic approaches is presented in Fig.~\ref{fig:freq}. In this figure, the observed frequencies $\bar{\Omega}_X$ and $\bar{\Omega}_Y$ are plotted as a function of the force $f_X$, with the other parameters fixed. Here and from hereon, we consider two types of coupling between the oscillators: the symmetric coupling $a_X=a_Y=0.5$ and the asymmetric coupling $a_X=0.9$, $a_Y=0$. In the latter case, the dynamics of the $Y$ oscillator is not affected by the $X$ oscillator. In all plots in our paper, we also always consider the same parameters as in Fig.~\ref{fig:freq}.

In the deterministic limit $N \rightarrow \infty$, the model exhibits distinct behaviors for $f_X$ below and above the critical value $f_X^*$ denoted by vertical gray lines. For $f_X<f_X^*$, corresponding to the unsynchronized state, the frequencies $\bar{\Omega}_X$ and $\bar{\Omega}_Y$ form two separate branches. For $f_X \geq f_X^*$, corresponding to a synchronized state, these branches join to a single branch with a common frequency $\bar{\Omega}$. Consequently, the observed frequencies $\bar{\Omega}_X$ and $\bar{\Omega}_Y$ are continuous but nonanalytic at $f_X=f_X^*$, showing that the transition from unsynchronized to synchronized state has the character of a continuous nonequilibrium phase transition. We also note that in the asymmetric case, the observed frequency $\bar{\Omega}_Y$ does not depend on $f_X$ and is equal to the intrinsic frequency $\Omega_Y$.

For finite $N$, the results are very close to the predictions of the deterministic approach even for relatively small $N=40$, provided that $f_X$ is sufficiently far from the phase transition point $f_X^*$. Close to the phase transition point, one may observe a substantial deviation of the results for small $N=40$ from the deterministic results, as the phase transition is blurred. However, this deviation is reduced with increasing system size $N$, demonstrating the asymptotic validity of the deterministic approach.

\subsection{Critical behavior} \label{subsec:critexp}
Let us now explore the behavior near the phase transition in closer detail, focusing first on the deterministic case. To that end, it is convenient to analyze the behavior of the frequency detuning $\vartheta$ defined by Eq.~\eqref{eq:freqdetuning}. We consider the situation where the system exhibits a phase transition due to a change of some parameter $\xi$ (e.g., $\xi=f_X$), so that the oscillators are unsynchronized (synchronized) for $\xi$ below (above) the critical value $\xi^*$. Then, for $\xi>\xi^*$ we have $\vartheta=0$, while for $\xi<\xi^*$ the frequency detuning is finite and given by Eq.~\eqref{eq:freqdetuning}. Close to $\xi^*$, one may apply the Taylor expansion
\begin{align}
\frac{K^2}{\omega^2}=1+g(\xi-\xi^*)+O\left[(\xi-\xi^*)^2\right],
\end{align}
where
\begin{align}
g \equiv \left. \frac{d}{d\xi}\frac{K^2}{\omega^2} \right \vert_{\xi=\xi^*} \,.
\end{align}
As a result, for $\xi$ below but close to $\xi^*$, $\vartheta$ exhibits universal critical behavior with a critical exponent $1/2$,
\begin{align} \label{eq:critbeh}
\vartheta \sim \omega^* \sqrt{|g|} \times (\xi^*-\xi)^{1/2} \,,   
\end{align}
where $\omega^*=\omega \vert_{\xi=\xi^*}$. Consequently, the response $d\vartheta/d\xi$ becomes divergent at the critical point $\xi^*$, which is typical for second-order phase transitions.

\subsection{Finite-size scaling of frequency detuning} \label{subsec:scalingdetuning}
We now note that frequency detuning $\vartheta\equiv\bar{\Omega}_Y-\bar{\Omega}_X$  remains well defined for finite $N$ using the stochastic frequency definitions~\eqref{eq:stochfreq}. In this case, the behavior of $\vartheta$ is no longer nonanalytic. On the other hand, for large $N$, it can be shown to exhibit universal scaling with $N$ close to the phase transition point $\xi^*$. To show that, we employ the van Kampen's expansion~\cite{van1992stochastic} of the reduced master equation~\eqref{eq:masteq-eff} for probabilities of the phase-difference states. For large $N$, the dominant terms produce a Langevin equation for the dynamics of the phase difference $\varphi\equiv \theta_Y-\theta_X$,
\begin{align} \label{eq:langevin}
d_t \varphi =F(\varphi)+\sqrt{2 D(\varphi)/N} \eta(t) \,,
\end{align}
where $F(\varphi)$ is the deterministic drift term of the Adler equation [Eq.~\eqref{eq:adler}], $\eta(t)$ is a zero-mean Gaussian white noise with correlation $\langle \eta(t) \eta(t') \rangle= \delta(t-t')$, and
\begin{align} \nonumber\label{eq:diffusion}
&D(\varphi)  = \frac{4\pi^2}{2} \sum_{\alpha \in\{X, Y \}} \left[w_+^{\alpha}(\varphi)+w_-^{\alpha}(\varphi) \right] \\ &=2 \pi^2 \left\{ \Gamma_X \left[1+a_X \sin(\varphi) \right]+\Gamma_Y \left[1-a_Y \sin(\varphi) \right] \right \} \,,
\end{align}
is the diffusion coefficient, where
\begin{subequations} \label{eq:intrates}
\begin{align}
w^X_\pm(\varphi) &= \lim_{N \rightarrow \infty} N^{-1} W_{i \pm 1,i}^X=\Gamma_X \frac{1+a_X \sin(\varphi)}{1+e^{\pm \beta f_X}} \,, \\
w^Y_\pm(\varphi) &= \lim_{N \rightarrow \infty} N^{-1} W_{i \pm 1,i}^Y=\Gamma_Y \frac{1-a_Y \sin(\varphi)}{1+e^{\mp \beta f_X}} \,,
\end{align}
\end{subequations}
are intensive transition rates. The frequency detuning then corresponds to the average phase difference velocity, $\vartheta=\langle d_t \varphi \rangle$.

We now note that Eq.~\eqref{eq:langevin} corresponds to diffusion in an effective tilted periodic potential
\begin{align} \label{eq:tilted-potential}
V(\varphi)=-\int_0^\varphi F(\varphi) d\varphi=K \left[1-\cos(\varphi) \right]-\omega \varphi \,,
\end{align}
so that the drift term corresponds to the gradient of that potential, $F(\varphi)=-d_\varphi F(\varphi)$. We also focus on a symmetric case with $\Gamma_X=\Gamma_Y$ and $a_X=a_Y$. Then the diffusion coefficient is homogeneous in space,
\begin{align} \label{eq:diffhom}
D(\varphi)=D_0  \equiv 2 \pi^2 \left(\Gamma_X+\Gamma_Y \right) \,.
\end{align}
This enables us to employ the theory developed in Refs.~\cite{reiman2001giant,reiman2002diffusion} to describe diffusion in tilted periodic potentials. Using this approach, close to $\xi^*$, the frequency detuning can be shown to scale as
\begin{align} \label{eq:scalingdetuning}
\vartheta \overset{N \rightarrow \infty}{=} N^{-1/3} \times 2 \pi D_0^{1/3} \mu^{2/3} U(\gamma) \,,
\end{align}
where
\begin{subequations} \label{eq:mugamma}
\begin{align}
\mu &\equiv - \frac{1}{6} \left. \frac{d^3}{d \varphi^3} V(\varphi) \right \vert_{\varphi=\varphi_c, \xi=\xi^*} >0 \,, \\
\gamma & \equiv \left. \frac{N^{2/3} (\xi-\xi^*)}{D_0^{2/3} \mu^{1/3}} \times \frac{d F(\varphi)}{d \xi} \right \vert_{\varphi=\varphi_c, \xi=\xi^*} \,,
\end{align}
\end{subequations}
with $\varphi_c=\pi \sgn(K/\omega)/2 \vert_{\xi=\xi^*}$. The function $U(\gamma)$ is the universal scaling function of the nondimensional parameter $\gamma$ (i.e., the rescaled difference $\xi-\xi^*$) expressed as
\begin{figure}
    \centering
    \includegraphics[width=0.95\linewidth]{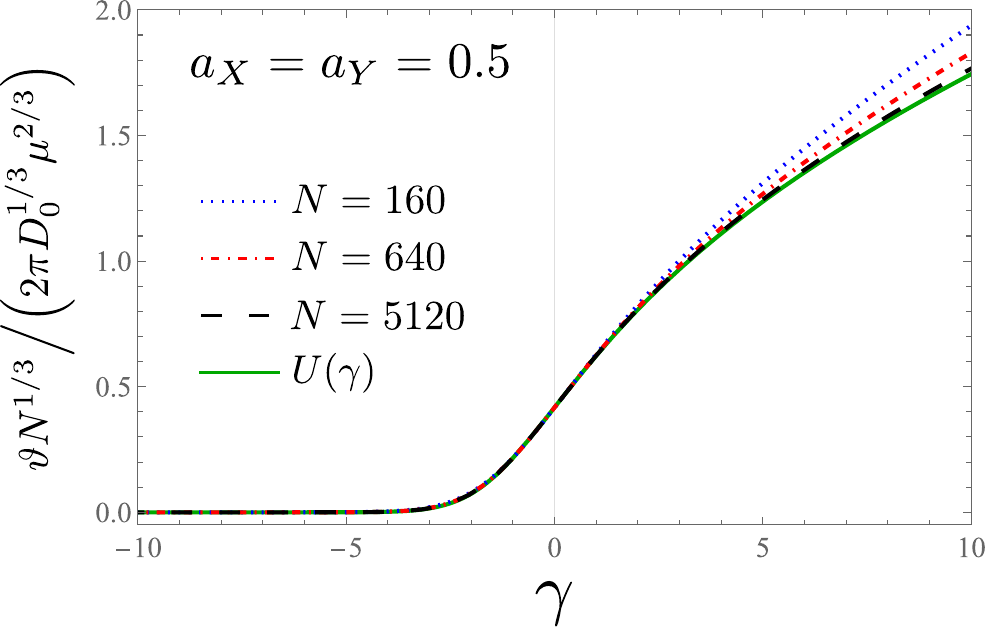}
    \caption{Demonstration of the validity of the scaling formula~\eqref{eq:scalingdetuning} for $\xi=f_X$. Rescaled frequency detuning $\vartheta N^{1/3}/(2\pi D_0^{1/3} \mu^{2/3})$ gradually converges with $N$ to a universal scaling function $U(\gamma)$. We consider symmetric coupling $a_X=a_Y=0.5$ and other parameters as in Fig.~\ref{fig:freq}:  $f_Y=2$, $\Gamma_X=\Gamma_Y=1$, $\beta=1$.}
    \label{fig:scalingform-detuning}
\end{figure}
\begin{align} \label{eq:scalfundetuning}
U(\gamma) = \frac{\sqrt[3]{9}}{\pi^2 \left[\text{Ai}(-\gamma/\sqrt[3]{3})^2+\text{Bi}(-\gamma/\sqrt[3]{3})^2 \right]} \,,
\end{align}
where $\text{Ai}$ and $\text{Bi}$ are the Airy functions of the first and second kind, respectively. In Fig.~\ref{fig:scalingform-detuning} we illustrate the validity of the scaling formula~\eqref{eq:scalingdetuning} for $\xi=f_X$. As shown there, the rescaled frequency detuning $\vartheta N^{1/3}/(2\pi D_0^{1/3} \mu^{2/3})$ gradually converges with $N$ to a universal scaling function $U(\gamma)$. 

\subsubsection{Scaling of maximum response}
Equation~\eqref{eq:scalingdetuning} further allows us to quantify the scaling of the response $d\vartheta/d\xi$ close to $\xi^*$. As discussed in Sec.~\ref{subsec:critexp}, in the deterministic limit, this response is divergent at $\xi=\xi^*$. For finite $N$, the response is finite. However, the emergence of divergent behavior manifests itself with the growth of the response magnitude as $N^{1/3}$, 
\begin{align}
\frac{d \vartheta}{d \xi} \overset{N \rightarrow \infty}{=} \left. N^{1/3} \times 2\pi \left(\frac{\mu}{D_0} \right)^{1/3} \frac{dU(\gamma)}{d \gamma} \frac{d F(\varphi)}{d \xi} \right \vert_{\varphi=\varphi_c, \xi=\xi^*} \,.
\end{align}
We now seek the maximum absolute response, $\max_\xi |d\vartheta/d\xi|$, which is obtained by maximizing $dU(\gamma)/d\gamma$ over $\gamma$. This maximum is located at $\gamma=0$, so that the response is maximized at the critical point $\xi=\xi^*$. The maximum response scales thus as
\begin{figure}
    \centering
    \includegraphics[width=0.95\linewidth]{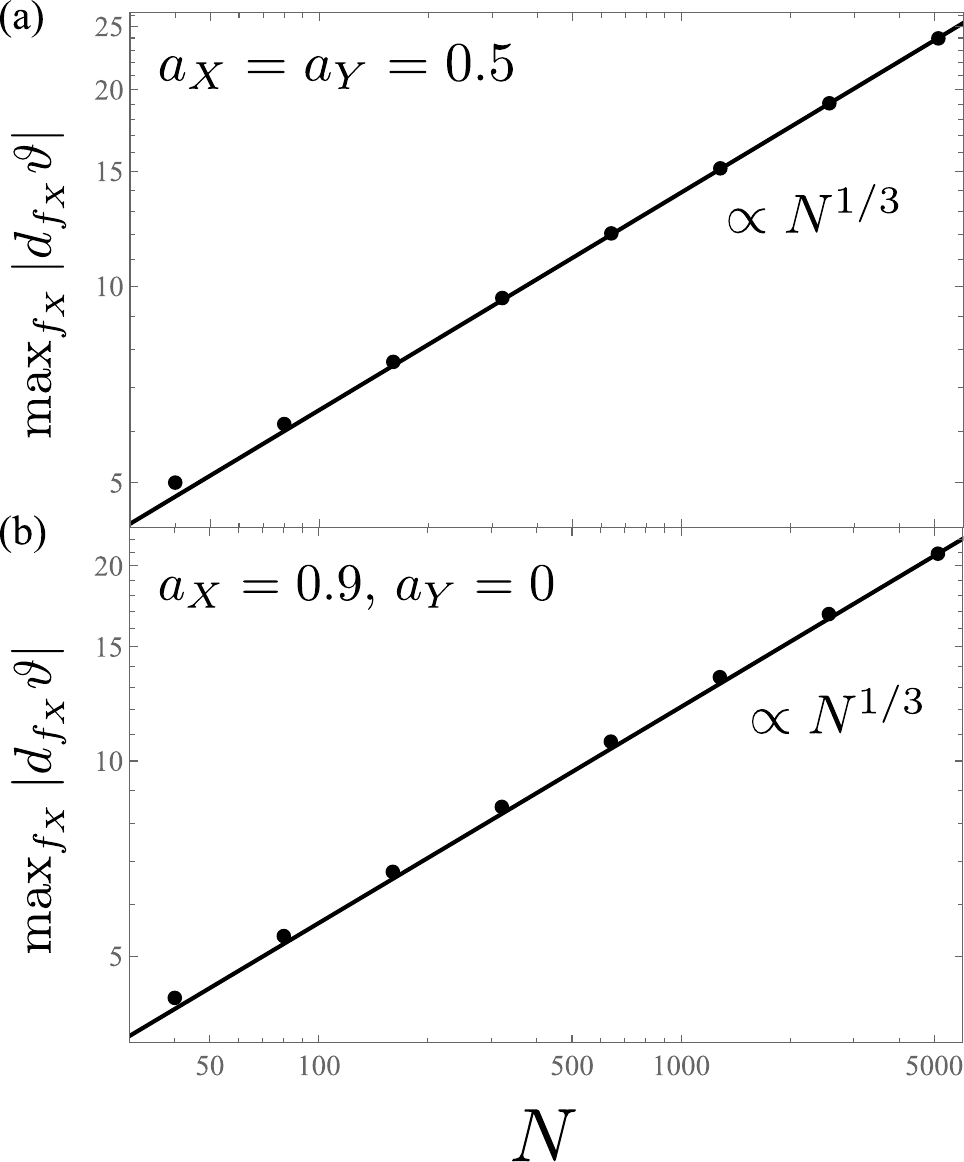}
    \caption{Demonstration of the $\propto N^{1/3}$ scaling of the maximum response of the frequency detuning, $\max_{f_X} |d_{f_X} \vartheta|$. It is plotted in the log-log scale for (a) the symmetric coupling $a_X=a_Y=0.5$ and (b) the asymmetric coupling $a_X=0.9$, $a_Y=0$. The points represent the master equation results. The black solid line in (a) represents Eq.~\eqref{eq:maxresponse}, while in (b) it represents $\propto N^{1/3}$ scaling fitted to cross the point for $N=5120$. Other parameters as in Fig.~\ref{fig:freq}:  $f_Y=2$, $\Gamma_X=\Gamma_Y=1$, $\beta=1$.}
    \label{fig:detuning-response}
\end{figure}
\begin{align} \label{eq:maxresponse}
\max_\xi \left \vert \frac{d \vartheta}{d \xi} \right\vert \overset{N \rightarrow \infty}{=} N^{1/3} \times A \left(\frac{\mu}{D_0} \right)^{1/3} \left \vert \frac{d F(\varphi)}{d \xi} \right \vert_{\varphi=\varphi_c, \xi=\xi^*} \,,
\end{align}
with $A \equiv \max_\gamma d_\gamma U(\gamma)=d_\gamma U(\gamma)|_{\gamma=0}=9 \Gamma(2/3)^3/[2\pi\Gamma(1/3)] \approx 1.33$, where $\Gamma$ is the gamma function. The validity of this expression for $\xi=f_X$ is demonstrated in Fig.~\ref{fig:detuning-response}~(a). In Fig.~\ref{fig:detuning-response}~(b) we further show that the same scaling $\max_{f_X} |d_{f_X} \vartheta| \propto N^{1/3}$ is also observed (at least, approximately) for the asymmetric coupling $a_X=0.9$, $a_Y=0$, where the above derivation is not directly applicable. Thus, this scaling appears to be a universal feature of the synchronization transition in our model.

\section{Nonequilibrium thermodynamics} \label{sec:therm}
Now we go beyond the kinetic description of the system to characterize its nonequilibrium thermodynamics. To that end, we consider the intensive local entropy production rate defined as
\begin{align}
\dot{\sigma}_\alpha \equiv \beta \dot{Q}^\text{dis}_\alpha/N \,,
\end{align}
where $\dot{Q}^\text{dis}_\alpha$ is the energy dissipation rate of the oscillator $\alpha$. Here we employ the fact that the system is in the stationary state, so that the contribution to entropy production rate related to entropy change of the system vanishes, and only the contribution related to energy dissipation to the environment remains.\footnote{We note that the local entropy production rate can also alternatively be defined by taking into account the information flow between the subsystems, which is defined in Sec.~\ref{subsec:infflow}; see Ref.~\cite{horowitz2014thermodynamics} for details.} We now note that the oscillator $X$ ($Y$) dissipates energy $f_X$ ($f_Y$) during every jump $x \rightarrow x+1$ ($y \rightarrow y+1)$, while extracting the same value of energy from the environment during the opposite jump. Thus, the energy dissipation can be calculated as
\begin{subequations}
\begin{align}
\dot{Q}^\text{dis}_X &\equiv f_X \sum_{\pm} \sum_{x,y} \pm W^y_{x \pm 1,x} p_{xy} \,, \\ \dot{Q}^\text{dis}_Y &\equiv f_Y \sum_{\pm} \sum_{x,y} \pm W^{y \pm 1,y}_x p_{xy} \,.
\end{align}
\end{subequations}
Consequently, using Eq.~\eqref{eq:stochfreq}, the local entropy production rate is proportional to the observed frequency of the corresponding oscillator,
\begin{align} \label{eq:locentrproddet}
\dot{\sigma}_\alpha=\frac{\beta f_\alpha \bar{\Omega}_\alpha}{2\pi} \,.
\end{align}
We further consider the intensive global entropy production rate
\begin{align} \label{eq:globentrproddet}
\dot{\sigma} \equiv\dot{\sigma}_X+\dot{\sigma}_Y \,.
\end{align}
By virtue of the second law of thermodynamics, the latter quantity needs to be non-negative: $\dot{\sigma} \geq 0$. This is guaranteed by construction by the local detailed balance condition~\eqref{eq:locdetbal}~\cite{Seifert2012}.

We now recall that in the deterministic limit, the observed frequencies $\bar{\Omega}_\alpha$ behave continuously but nonanalytically at the synchronization phase transition. Consequently, since local entropy production rates $\dot{\sigma}_\alpha$ are proportional to observed frequencies, both local and global entropy production rates are also nonanalytic at the phase transition point. This is illustrated explicitly in Fig.~\ref{fig:entrprod} in the next section. Thus, the nonanalytic behavior of the entropy production rate serves as a witness of the nonequilibrium phase transition, as previously discussed in Ref.~\cite{noa2019entropy}.

\subsection{Thermodynamic behavior of the system: universalities versus nonuniversalities}

\subsubsection{Close-to-equilibrium behavior}
Now, we investigate whether the thermodynamic behavior of the system obeys some universal thermodynamic principles beyond the second law of thermodynamics $\dot{\sigma} \geq 0$. First, we ask whether our model is consistent with the established principles of nonequilibrium thermodynamics close to thermodynamic equilibrium (i.e., for small forces $f_\alpha$). In this regime, the observed frequencies $\bar{\Omega}_\alpha$ should be linear in applied thermodynamic forces, $\bar{\Omega}_\alpha=\sum_{\alpha'}L_{\alpha \alpha'} f_{\alpha'}$, with the linear-response coefficients $L_{\alpha \alpha'}$ satisfying Onsager reciprocity relations $L_{\alpha \alpha'}=L_{\alpha' \alpha}$. At the level of stochastic description obeying the local detailed balance condition~\eqref{eq:locdetbal}, the Onsager relations are satisfied by construction~\cite{Seifert2012,schnakenberg1976network,forastiere2022linear}. However, an apparent inconsistency appears for the deterministic description. Then, for small forces $f_\alpha$, we can expand the intrinsic frequencies $\Omega_\alpha$ and couplings $K_\alpha$ linearly in terms of forces,
\begin{subequations} \label{eq:smallforceexp}
\begin{align}
\Omega_\alpha &=\pi \beta \Gamma_\alpha f_\alpha+O(f_\alpha^2) \,, \\ K_\alpha &=\pi \beta a_\alpha \Gamma_\alpha f_\alpha+O(f_\alpha^2) \,.
\end{align}
\end{subequations}
Inserting these expressions into Eqs.~\eqref{eq:freqsync} and~\eqref{eq:solomegaxdet}, we find that even for small forces $f_\alpha$ the observed frequencies $\bar{\Omega}_\alpha$ are nonlinear functions of forces, i.e., they cannot be expressed as $\bar{\Omega}_\alpha=\sum_{\alpha'}L_{\alpha \alpha'} f_{\alpha'}$. Furthermore, the deterministic description appears to violate the postulate of Ref.~\cite{nicolis1970thermodynamic} that close to equilibrium all macroscopic states of the system are continuous extensions of its equilibrium states (which form the so-called ``thermodynamic branch'' of macroscopic states). In our model, independent of the system parameters, the equilibrium state corresponds to a uniform distribution of the equally probable phase difference states, $p_i=1/N$. In contrast, at the level of deterministic description, even when we apply the small-force expansion~\eqref{eq:smallforceexp}, the system possesses two distinct macroscopic states: the synchronized state for $|\Gamma_Y f_Y-\Gamma_X f_X| <|a_Y \Gamma_Y f_Y+a_X\Gamma_X f_X|$ and the unsynchronized state otherwise. Only the unsynchronized state can be regarded as a continuous extension of the equilibrium state, as it corresponds to a smooth distribution of phase difference states. In contrast, the synchronized state is not a continuous extension of the equilibrium state, as it corresponds to a well-defined phase difference $\varphi=2\pi i/N$.

This paradox can be resolved by applying a more rigorous definition of the close-to-equilibrium regime: following Refs.~\cite{schnakenberg1976network,andrieux2007fluctuation,forastiere2022linear}, we state that the system is close to equilibrium when the cycle affinities $\mathcal{A}_\alpha=N \beta f_\alpha$ are small. Here, the affinity $\mathcal{A}_\alpha$ corresponds to the entropy produced by the oscillator $\alpha$ when its phase $\theta_\alpha$ increases by $2\pi$. Applying this definition, we find that close to equilibrium the drift term $F(\varphi)$ of the Langevin equation~\eqref{eq:langevin} is of the order $O(1/N)$, which is of the same order as the noise term. Consequently, for small $\mathcal{A}_\alpha$, the deterministic description is not applicable and the stochastic description is required. The latter provides a unique steady state, which is a continuous extension of the equilibrium state. 

In summary, when the affinities $\mathcal{A}_\alpha$ are small so that the forces $f_\alpha$ are of the order $O(1/N)$, the system exhibits a linear response behavior that is not captured by the deterministic description. On the other hand, whenever the forces $f_\alpha$ are of the order $O(1)$, the deterministic description is applicable, and the system response to forces is nonlinear. This further means that the range of forces $f_\alpha$ where the linear-response regime is applicable shrinks as $1/N$. This unusual behavior is a consequence of the peculiar nature of the system considered, where the deterministic drift $F(\varphi)$ vanishes at equilibrium, so that the system does not have any stable fixed point. It contrasts with situations more typical for macroscopic systems, where the deterministic drift does not vanish at equilibrium and the system exhibits stable fixed points, so that the linear-response regime is well defined also at the level of deterministic description~\cite{FalascoReview}.

\subsubsection{No extremum dissipation principle}
Second, we raise the question of whether synchronization universally affects energy dissipation. Previous studies on this topic have identified models in which synchronization always enhances~\cite{zhang2020energy, gusilain2024discontinuous} or reduces~\cite{izumida2016energetics,
herpich2018collective,meibohm2024minimum,meibohm2024small,gopal2025dissipation} dissipation, independent of their parameters. We found that in our system there is no such universality: synchronization may either increase or decrease dissipation, depending on the system parameters. To show that, we define two measures of the change in dissipation due to synchronization. One is more relevant for synchronized state and is defined as
\begin{align}
\Delta \dot{\sigma}_\text{sf} \equiv \dot{\sigma}_\text{sync}-\dot{\sigma}_\text{free} \,,
\end{align}
where $\dot{\sigma}_\text{sync} \equiv \beta (f_X+f_Y) \bar{\Omega}/(2\pi)$ is the actual entropy production rate of the synchronized state, and $\dot{\sigma}_\text{free} \equiv \beta (f_X \Omega_X+f_Y \Omega_Y)/(2\pi)$ is the entropy production rate for the reference system of uncoupled oscillators. The second definition is more relevant for the unsynchronized state and is defined as
\begin{align}
\Delta \dot{\sigma}_\text{su}\equiv \dot{\sigma}_\text{sync}-\dot{\sigma}_\text{unsync} \,,
\end{align}
where $\dot{\sigma}_\text{unsync} \equiv \beta(f_X \bar{\Omega}_X+f_Y \bar{\Omega}_Y)/(2\pi)$ is the actual entropy production rate of unsynchronized oscillators, and $\dot{\sigma}_\text{sync} \equiv (f_X+f_Y) \bar{\Omega}$, with $\bar{\Omega}$ given by Eq.~\eqref{eq:freqsync}, now corresponds to the entropy production rate for the reference synchronized state. Using previous results, both quantities can be expressed using compact expressions
\begin{subequations}
    \begin{align}
\Delta \dot{\sigma}_\text{sf} &=\frac{\beta \omega}{2\pi} \frac{K_X f_X-K_Y f_Y}{K} \,, \\ \label{eq:dsigmasu}
\Delta \dot{\sigma}_\text{su} &=\frac{\beta \vartheta}{2\pi} \frac{K_X f_X-K_Y f_Y}{K} \,,
\end{align}
\end{subequations}
which differ only by the factor $\vartheta/\omega=\sqrt{1-K^2/\omega^2}$.

We now find that only for identical oscillators $\Gamma_X=\Gamma_Y=\Gamma$ and reciprocal coupling $K_X=K_Y=K$ one can observe certain universality, namely, synchronization always reduces the dissipation:
\begin{subequations}
\begin{align}
\Delta \dot{\sigma}_\text{sf} &=-\frac{\beta \Gamma}{2} \left(\tanh \frac{\beta f_Y}{2}-\tanh \frac{\beta f_X}{2} \right)(f_Y-f_X)  \leq 0 \,, \\ \nonumber
\Delta \dot{\sigma}_\text{su} &=-\frac{\beta \Gamma}{2} \sqrt{1-\frac{K^2}{\omega^2}} \left(\tanh \frac{\beta f_Y}{2}-\tanh \frac{\beta f_X}{2} \right)(f_Y-f_X)  \\ &\leq 0 \,.
\end{align}
\end{subequations}
We note that this situation is comparable to the odd-coupling case from Ref.~\cite{izumida2016energetics}, for which the reduction of dissipation by synchronization has been reported. 
Apart from this specific regime, synchronization can either reduce or enhance dissipation, i.e., the sign of $\Delta \dot{\sigma}_\text{sf}$ and $\Delta \dot{\sigma}_\text{su} $ can be either negative or positive. In particular, for a fully asymmetric coupling $K_Y=0$, we obtain
\begin{subequations}
\begin{align}
\Delta \dot{\sigma}_\text{sf} &=\beta f_X\omega \,, \\
\Delta \dot{\sigma}_\text{su} &=\beta f_X \omega \sqrt{1-K^2/\omega^2} \,,
\end{align}
\end{subequations}
so that the sign of $\Delta \dot{\sigma}_\text{sf}$ and $\Delta \dot{\sigma}_\text{su} $ is determined by the sign of $f_X \omega$. Since this sign is arbitrary, there is no extremum dissipation principle. We note that a similar situation has been observed in Ref.~\cite{izumida2016energetics} for the case where the coupling between oscillators contains an even contribution, i.e., Eq.~\eqref{eq:detphasedyn} is replaced with $d_t \theta_\alpha=\Omega_\alpha-K C(\theta_\alpha-\theta_{\alpha'})$, where $C(\varphi) \neq -C(-\varphi)$. In contrast, in our case this occurs when the coupling is odd, since in Eq.~\eqref{eq:detphasedyn} we have $\sin(\varphi)=-\sin(-\varphi)$, but nonreciprocal.

\begin{figure}
    \centering
    \includegraphics[width=0.95\linewidth]{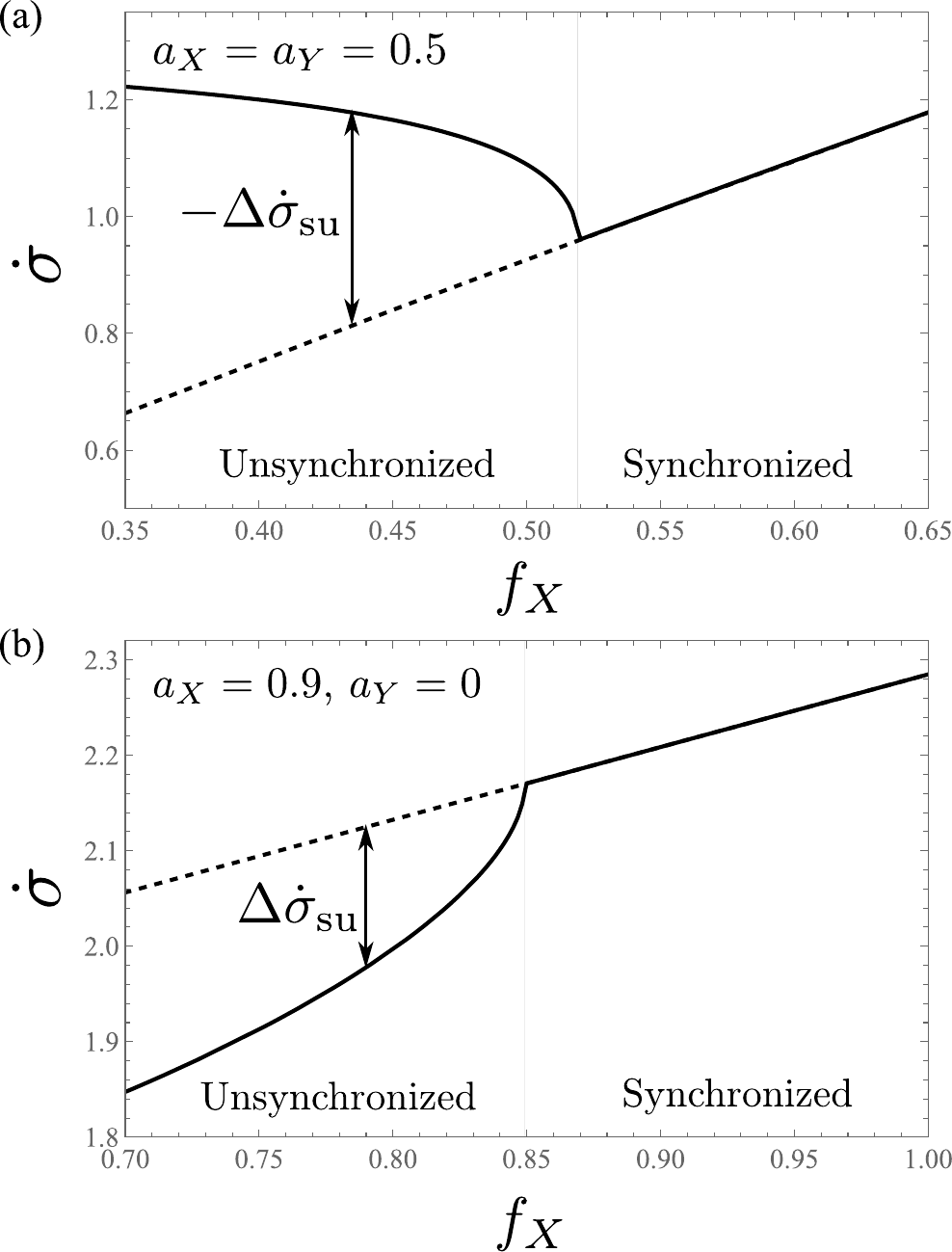}
    \caption{Intensive global entropy production rate $\dot{\sigma}$, calculated in the deterministic limit, as a function of $f_X$ for (a) the symmetric coupling $a_X=a_Y=0.5$ and (b) the asymmetric coupling $a_X=0.9$, $a_Y=0$. The vertical gray lines denote the phase transition from unsynchronized (left) to synchronized (right) state. The dashed line represents the extrapolation of trend from the synchronized state. Other parameters are as in Fig.~\ref{fig:freq} ($f_Y=2$, $\Gamma_X=\Gamma_Y=1$, $\beta=1$), but the results are plotted for a smaller range of $f_X$ for better visibility.}
    \label{fig:entrprod}
\end{figure}
To further illustrate this observation, and the physical meaning of $\Delta \dot{\sigma}_\text{su}$, in Fig.~\ref{fig:entrprod} we plot the behavior of the entropy production rate near the phase transition point. Then $\Delta \dot{\sigma}_\text{su}$ corresponds to a difference between the actual entropy production rate (solid line) and the extrapolation of the trend from the synchronized state (dashed line). As shown, this difference may be either negative or positive, illustrating the lack of a minimum or maximum dissipation principle in our model.  We also note that the quantity $\Delta \dot{\sigma}_{\text{su}}$ is proportional to $\vartheta$ [see Eq.~\eqref{eq:dsigmasu}]. Thus, close to the synchronization transition, it follows the same critical behavior with the critical exponent $1/2$ [see Eq.~\eqref{eq:critbeh}]. A similar critical behavior has been observed for the even-coupling contribution to energy dissipation in Ref.~\cite{izumida2016energetics}.

\subsubsection{No Maxwell demons in the deterministic limit} \label{subsec:nodemons}
Finally, we ask whether the considered system can work as an autonomous Maxwell demon in the deterministic limit. This name refers to bipartite stochastic systems in which one of the subsystems can continuously reduce the entropy of its environment, $\dot{\sigma}_\alpha <0$, due to autonomous control by the other subsystem~\cite{horowitz2014thermodynamics,strasberg2013thermodynamics,koski2015on}. Previous studies of electronic systems~\cite{freitas2022maxwell,freitas2023information} and chemical systems~\cite{bilancioni2023chemical} suggest that Maxwell demon operation is not possible in the deterministic limit of stochastic systems. We found that the same principle applies to the setup considered, as the local entropy production rates $\dot{\sigma}_\alpha$ are always nonnegative. To show that, we use the fact that the observed frequency $\bar{\Omega}_\alpha$ has the same sign as the intrinsic frequency $\Omega_\alpha$ (see Sec.~\ref{subsec:nochanges}), and thus the same sign as the force $f_\alpha$ [see Eq.~\eqref{eq:defomega}]. Then, from  Eq.~\eqref{eq:locentrproddet}, we have
\begin{align}
\sgn(\dot{\sigma}_\alpha)=\sgn(f_\alpha) \sgn(\bar{\Omega}_\alpha)=\sgn(\Omega_\alpha) \sgn(\bar{\Omega}_\alpha)=1 \,,
\end{align}
so that $\dot{\sigma}_\alpha \geq 0$. However, we note that Maxwell demon behavior can be observed in our model for very small systems, such as $N=3$. For example, for parameters such as in Fig.~\ref{fig:freq}~(a), it is observed in a small region $0.09 \lessapprox f_X<0$.

\section{Fluctuations} \label{sec:fluct}

\subsection{Phase fluctuations: definitions}
Thus far we have focused on the behavior of the average quantities, such as observed frequency or entropy production rate. We now aim to characterize the stochastic behavior of the phase of oscillators. To do this, we first define the stochastic counting variable $n_X(t)$ [$n_Y(t)$] as the number of jumps $x \rightarrow x+1$ ($y \rightarrow y+1$) minus the number of opposite jumps $x \rightarrow x-1$ ($y \rightarrow y-1$) in the time interval $[0,t]$. We further define the stochastic phase of the oscillator $\alpha$ as
\begin{align} \label{eq:stochphase}
\theta_\alpha(t) \equiv 2 \pi n_\alpha(t) /N\,.
\end{align}
The observed frequency of the oscillator $\alpha$, defined by Eq.~\eqref{eq:stochfreq}, corresponds then to the average phase velocity,
\begin{align}
\bar{\Omega}_\alpha = \lim_{t \rightarrow \infty} t^{-1} \left \langle \theta_\alpha(t) \right \rangle \,,
\end{align}
where $\langle \cdot \rangle$ denotes the average over the ensemble of stochastic system trajectories. 

We then characterize the phase fluctuations by means of a scaled covariance of two phases,
\begin{align} \label{eq:covariance-def}
\llangle \theta_\alpha,\theta_{\alpha'} \rrangle \equiv \lim_{t \rightarrow \infty} \frac{N}{t} \left \langle \Delta \theta_\alpha(t) \Delta \theta_{\alpha'}(t) \right \rangle \,,
\end{align}
where $\Delta \theta_\alpha(t)=\theta_\alpha(t)-\left \langle \theta_\alpha(t) \right \rangle $. Scaling by $N$ ensures that this quantity does not vanish for $N \rightarrow \infty$. In particular, the variance of a single phase is defined as its autocovariance,
\begin{align}
\llangle \theta_\alpha \rrangle \equiv \llangle \theta_\alpha,\theta_{\alpha} \rrangle \,.
\end{align}

\subsection{Methods}
To determine the above quantities, we employ the spectral approach to full counting statistics that has been recently thoroughly reviewed in Ref.~\cite{landi2023current}; see also Refs.~\cite{wachtel2015fluctuating,aslyamov2024nonequilibriumfluctuation} for other methods. Within this approach, we define the counting field dependent rate matrix $\mathbb{W}^\psi$ with the same diagonal elements as the matrix $\mathbb{W}$ defined in Sec.~\ref{subsec:reduction}. The off-diagonal elements are defined as
\begin{align}
W^\psi_{i \pm 1,i}= W_{i \pm 1,i}^{X} e^{\mp 2\pi \chi_X}+W_{i \pm 1,i}^{Y} e^{\pm 2\pi \chi_Y} \,.
\end{align}
The average phase velocity can be then calculated as
\begin{align}
\bar{\Omega}_\alpha =\boldsymbol{1}^\intercal \mathbb{J}^{(1)}_\alpha \boldsymbol{p}/N \,,
\end{align}
where
\begin{align}
\mathbb{J}^{(k)}_\alpha \equiv \left( \frac{\partial^k \mathbb{W}^\psi}{\partial \chi_\alpha^k} \right)_{\chi_X=0,\chi_Y=0} \,,
\end{align}
and $\boldsymbol{1} \equiv (\ldots,1,\ldots)^\intercal$ is the column vector of ones with the length $N$. The phase covariance can be calculated as
\begin{align} \nonumber
\llangle \theta_\alpha,\theta_{\alpha'} \rrangle =\ &\delta_{\alpha \alpha'} \boldsymbol{1}^\intercal \mathbb{J}_\alpha^{(2)} \boldsymbol{p}/N - \boldsymbol{1}^\intercal \mathbb{J}_\alpha^{(1)} \mathbb{W}^D \mathbb{J}_{\alpha'}^{(1)} \boldsymbol{p}/N \\ & - \boldsymbol{1}^\intercal \mathbb{J}_{\alpha'}^{(1)} \mathbb{W}^D \mathbb{J}_{\alpha}^{(1)} \boldsymbol{p}/N \,,
\end{align}
where $\mathbb{W}^D$ is the Drazin inverse~\cite{drazin1958pseudo} of the matrix $\mathbb{W}$, a unique solution of the equation $\mathbb{W} \mathbb{W}^D=\mathds{1}-\boldsymbol{p} \boldsymbol{1}^\intercal$; see Refs.~\cite{crook2018drazin,landi2023current} for its properties and applications to characterize fluctuations.

\subsection{Fluctuations of phase difference}
Before considering covariances $\llangle \theta_\alpha,\theta_{\alpha'} \rrangle$ on their own, we first investigate the variance of the stochastic phase difference defined as
\begin{align} \label{eq:phasediffwar-def} 
\llangle \varphi \rrangle &\equiv \lim_{t \rightarrow \infty} \frac{N}{t} \left \langle \Delta \varphi(t)^2 \right \rangle \,,
\end{align}
where $\Delta \varphi(t) \equiv \Delta \theta_Y(t)- \Delta \theta_X(t)$.
It can be expressed in terms of covariances as
\begin{align} \label{eq:phasediffwar}
\llangle \varphi \rrangle =\llangle \theta_X \rrangle+\llangle \theta_Y \rrangle-2\llangle \theta_X,\theta_Y \rrangle \,.
\end{align}
The analysis of this quantity will enable us to gain certain analytic insight. To that end, we recall that for large $N$ the dynamics of the stochastic phase difference $\varphi$ can be modeled using the Langevin equation~\eqref{eq:langevin} for diffusion in a tilted periodic potential $V(\varphi)$. In the synchronized state, the phase difference relaxes to a minimum of that potential, which corresponds to the fixed point $\varphi^*$ given by Eq.~\eqref{eq:fixedpointadler}. For $N \rightarrow \infty$, the noise-induced escape rate from that minimum is exponentially suppressed with $N$, and thus $\lim_{N \rightarrow \infty} \llangle \varphi \rrangle=0$.

In the unsynchronized state, we can obtain an analytic expression for the variance that is asymptotically valid in the limit $N \rightarrow \infty$. To that end, we note two facts. First, using renewal theory, we obtain the relation $\llangle \varphi \rrangle/(2 \pi |\omega|)=N\langle \Delta \mathcal{T}^2 \rangle/T^2$~\cite{ptaszynski2018first}, where $\langle \Delta \mathcal{T}^2 \rangle$ is the variance of the first-passage time after which the phase difference increases by $2\pi$, and the period $T$ defined by Eq.~\eqref{eq:periodphasedif} corresponds to the average first-passage time. On the other hand, applying the Langevin equation~\eqref{eq:langevin} for large $N$, one obtains the relation $\langle \Delta \mathcal{T}^2 \rangle/T^2=|T|/(2 \pi^2 \tau_c)$~\cite{barroo2015fluctuating}, where $\tau_c$ is the correlation-time that describes the decay of two-time correlation functions. Combining these two results, we obtain $\llangle \varphi \rrangle/(2 \pi |\omega|)=N|T|/(2 \pi^2 \tau_c)$. Using $\omega=2\pi/T$ this yields
\begin{align} \label{eq:varviacortime}
\llangle \varphi \rrangle \overset{N \rightarrow \infty}{=} 2N/{\tau_c} \,.
\end{align}
Applying the Langevin equation~\eqref{eq:langevin} further, the correlation time can be determined as~\cite{remlein2022coherence}
\begin{align} \label{eq:cortimerem}
\tau_c \overset{N \rightarrow \infty}{=} \frac{N}{4 \pi ^2} \frac{\left[\int_0^{2\pi} d\varphi/F(\varphi) \right]^3}{ \int_0^{2\pi} d\varphi D(\varphi) /F(\varphi)^3} \,.
\end{align}
Combining Eqs.~\eqref{eq:varviacortime} and~\eqref{eq:cortimerem}, we obtain the desired asymptotic expression
\begin{figure}
    \centering
    \includegraphics[width=0.95\linewidth]{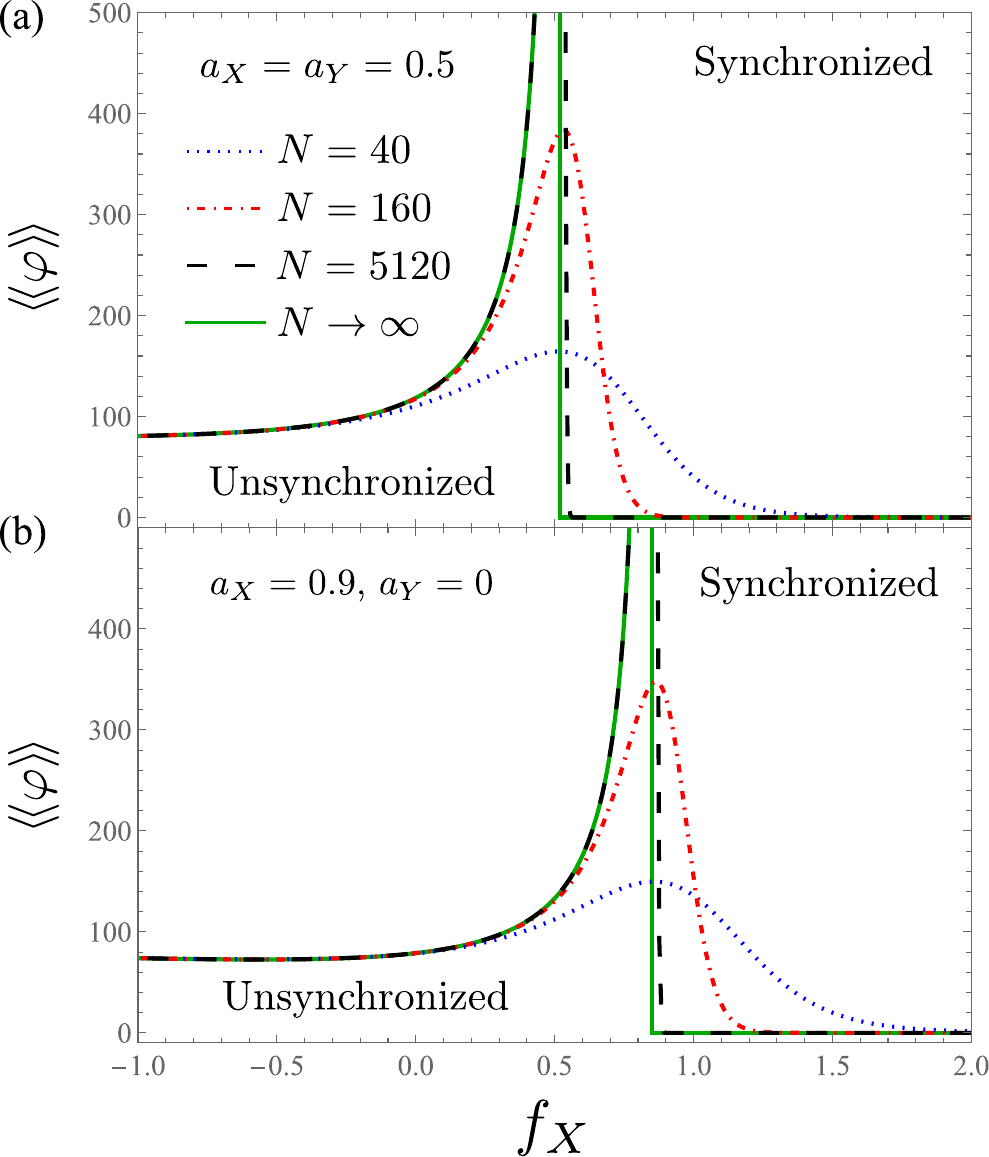}
    \caption{Phase difference variance $\llangle \varphi \rrangle$ as a function of $f_X$ for (a) the symmetric coupling $a_X=a_Y=0.5$ and (b) the asymmetric coupling $a_X=0.9$, $a_Y=0$. The results for $N \rightarrow \infty$ correspond to the asymptotic formula~\eqref{eq:asympvarphasedif}. They exhibit sharp jump at the transition from unsynchronized (left) to synchronized (right) state. Other parameters as in Fig.~\ref{fig:freq}:  $f_Y=2$, $\Gamma_X=\Gamma_Y=1$, $\beta=1$.}
    \label{fig:variance}
\end{figure}
\begin{align} \label{eq:asympvarphasedif}
&\lim_{N \rightarrow \infty} \llangle \varphi \rrangle= \\ \nonumber &\frac{2 \pi^2 \left[(\Gamma_X+\Gamma_Y)(2\omega^2+K^2)+3K\omega(a_X \Gamma_X-a_Y \Gamma_Y) \right]}{\omega^2-K^2} \,.
\end{align}

In Fig.~\ref{fig:variance} we compare the above analytic formula with the results for finite $N$. For $N \rightarrow \infty$, $\llangle \varphi \rrangle$ is discontinuous and nonanalytic at the phase transition point:  it is equal to 0 in the synchronized state, while diverges when approaching the phase transition point from the unsynchronized state. This divergence is related to the vanishing of the denominator in Eq.~\eqref{eq:asympvarphasedif}. In contrast, for finite $N$, the variance $\llangle \varphi \rrangle$ is finite and continuous.  Away from the phase transition point, quantitative agreement between the asymptotic formula~\eqref{eq:asympvarphasedif} and the finite-size results is already observed for relatively small $N=40$. Around the phase transition point, $\llangle \varphi \rrangle$ increases strongly with $N$, witnessing the gradual  emergence of infinite-size behavior.

The enhancement of fluctuations around the phase transition point can be explained using the mechanism of giant enhancement of diffusion described in Refs.~\cite{reiman2001giant,reiman2002diffusion}. It is related to the existence of a dynamical bottleneck around $\varphi =\pi \sgn(K/\omega)/2$, where the drift dynamics is very slow, $F(\varphi) \approx 0$. As a consequence, the time needed to cross this bottleneck becomes strongly influenced by the noise term of the Langevin equation~\eqref{eq:langevin}. This leads to a huge dispersion of phase difference growths $\Delta \varphi(t)$ for a statistical ensemble of trajectories subjected to different noise realizations, which enhances $\llangle \varphi \rrangle$. 

\subsubsection{Finite-size scaling}
Applying the theory presented in Sec.~\ref{subsec:scalingdetuning}, we can further provide some analytic insight about the scaling behavior of $\llangle \varphi \rrangle$ for large but finite $N$. We again focus on the symmetric case with $\Gamma_X=\Gamma_Y$ and $a_X=a_Y$. Then, the scaling of fluctuations close to the phase transition point can be expressed as~\cite{reiman2001giant,reiman2002diffusion}
\begin{align} \label{eq:scalingformvariance}
\llangle \varphi \rrangle  \overset{N \rightarrow \infty}{=} N^{2/3}\times 8\pi^2 D_0^{1/3} \mu^{2/3}G(\gamma) \,,
\end{align}
with $D_0$, $\mu$ and $\gamma$ defined by Eqs.~\eqref{eq:diffhom} and~\eqref{eq:mugamma}. The scaling function $G(\gamma)$ takes the form
\begin{align}
G(\gamma)=U(\gamma)^3 \int_{-\infty}^\infty dx\,S(x,\gamma)^2 S(-x,\gamma) \,,
\end{align}
with $U(\gamma)$ given by Eq.~\eqref{eq:scalfundetuning} and
\begin{align}
S(x,\gamma) \equiv \int_{0}^\infty dy\,e^{-x^3+(x-y)^3-\gamma y} \,.
\end{align}

\begin{figure}
    \centering
    \includegraphics[width=0.95\linewidth]{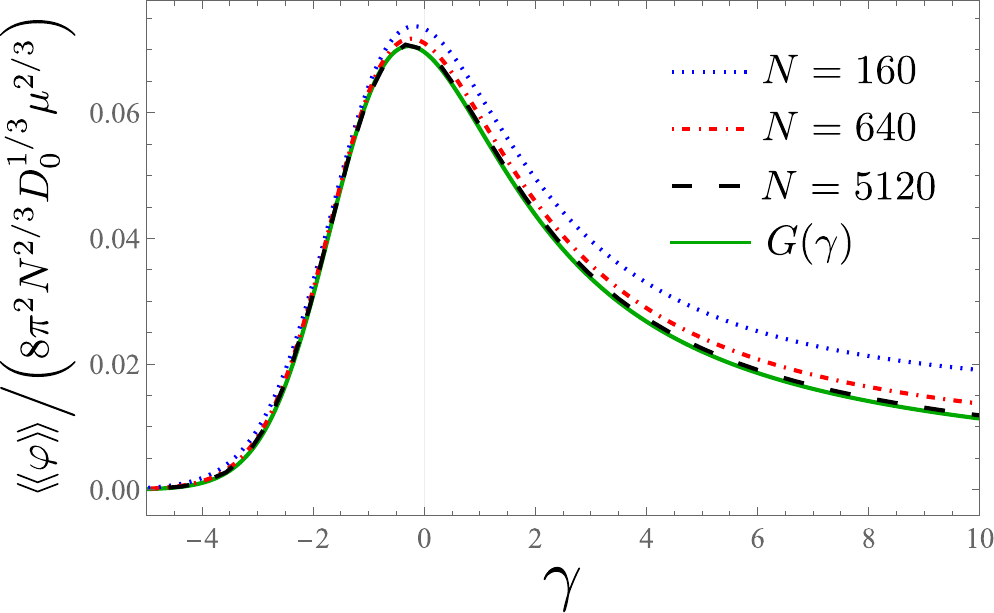}
    \caption{Demonstration of the validity of the scaling formula~\eqref{eq:scalingformvariance} for $\xi=f_X$. Rescaled variance $\llangle \varphi \rrangle/ (8\pi^2 N^{2/3} D_0^{1/3} \mu^{2/3})$ gradually converges with $N$ to a universal scaling function $G(\gamma)$. We consider symmetric coupling $a_X=a_Y=0.5$ and other parameters as in Fig.~\ref{fig:freq}:  $f_Y=2$, $\Gamma_X=\Gamma_Y=1$, $\beta=1$.}
    \label{fig:scalingform-variance}
\end{figure}

In Fig.~\ref{fig:scalingform-variance} we illustrate the validity of the scaling formula~\eqref{eq:scalingformvariance} for $\xi=f_X$. As shown in the figure, the rescaled variance $\llangle \varphi \rrangle/ (8\pi^2 N^{2/3} D_0^{1/3} \mu^{2/3})$ gradually converges with $N$ to a universal scaling function $G(\gamma)$. 

\begin{figure}
    \centering
    \includegraphics[width=0.95\linewidth]{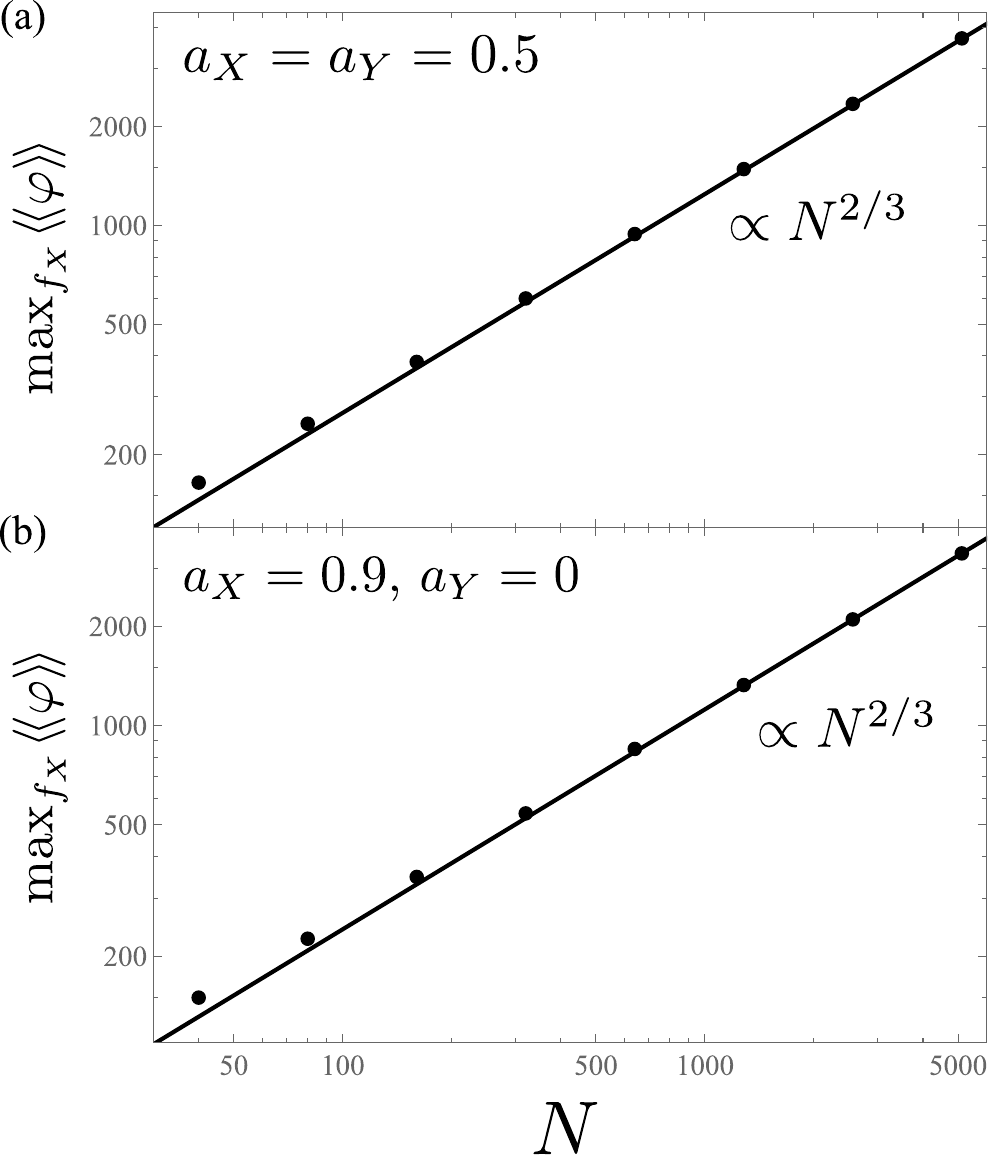}
    \caption{Demonstration of the $\propto N^{2/3}$ scaling of the maximum variance of phase difference, $\max_{f_X}\llangle \varphi \rrangle$. It is plotted in the log-log scale for (a) the symmetric coupling $a_X=a_Y=0.5$ and (b) the asymmetric coupling $a_X=0.9$, $a_Y=0$. The points represent the master equation results. The black solid line in (a) represents Eq.~\eqref{eq:maxvariance}, while in (b) it represents $\propto N^{2/3}$ scaling fitted to cross the point for $N=5120$. Other parameters as in Fig.~\ref{fig:freq}:  $f_Y=2$, $\Gamma_X=\Gamma_Y=1$, $\beta=1$.}
    \label{fig:scaling-variance}
\end{figure}

The formula~\eqref{eq:scalingformvariance} further allows us to determine the magnitude of the variance peak (i.e., its maximum as a function of $f_X$) that occurs close to the phase transition point (see Fig.~\ref{fig:variance}). For an arbitrary tuned parameter $\xi$, it can be calculated as
\begin{align} \label{eq:maxvariance} \nonumber
\max_\xi \llangle \varphi \rrangle  &\overset{N \rightarrow \infty}{=} N^{2/3}\times 8\pi^2 D_0^{1/3} \mu^{2/3} \times \max_\gamma G(\gamma) \\ 
&\approx N^{2/3} \times 5.56D_0^{1/3} \mu^{2/3} \,,
\end{align}
where we use $\max_\gamma G(\gamma) \approx 0.07$ at $\gamma \approx -0.28$. The validity of this expression for $\xi=f_X$ is demonstrated in Fig.~\ref{fig:scaling-variance}~(a). In Fig.~\ref{fig:scaling-variance}~(b) we further show that the same scaling $\max_{f_X} \llangle \varphi \rrangle \propto N^{2/3}$ is also observed for the asymmetric coupling $a_X=0.9$, $a_Y=0$. This suggests that such scaling is a universal feature of the synchronization transition in our model. We note that similar polynomial scaling of fluctuations has also been observed for other models of continuous phase transitions~\cite{nguyen2018phase,oberreiter2021stochastic,kewming2022diverging,remlein2024nonequilibrium,ptaszynski2024critical}.

\subsection{Phase covariances}
\begin{figure}
    \centering
    \includegraphics[width=0.95\linewidth]{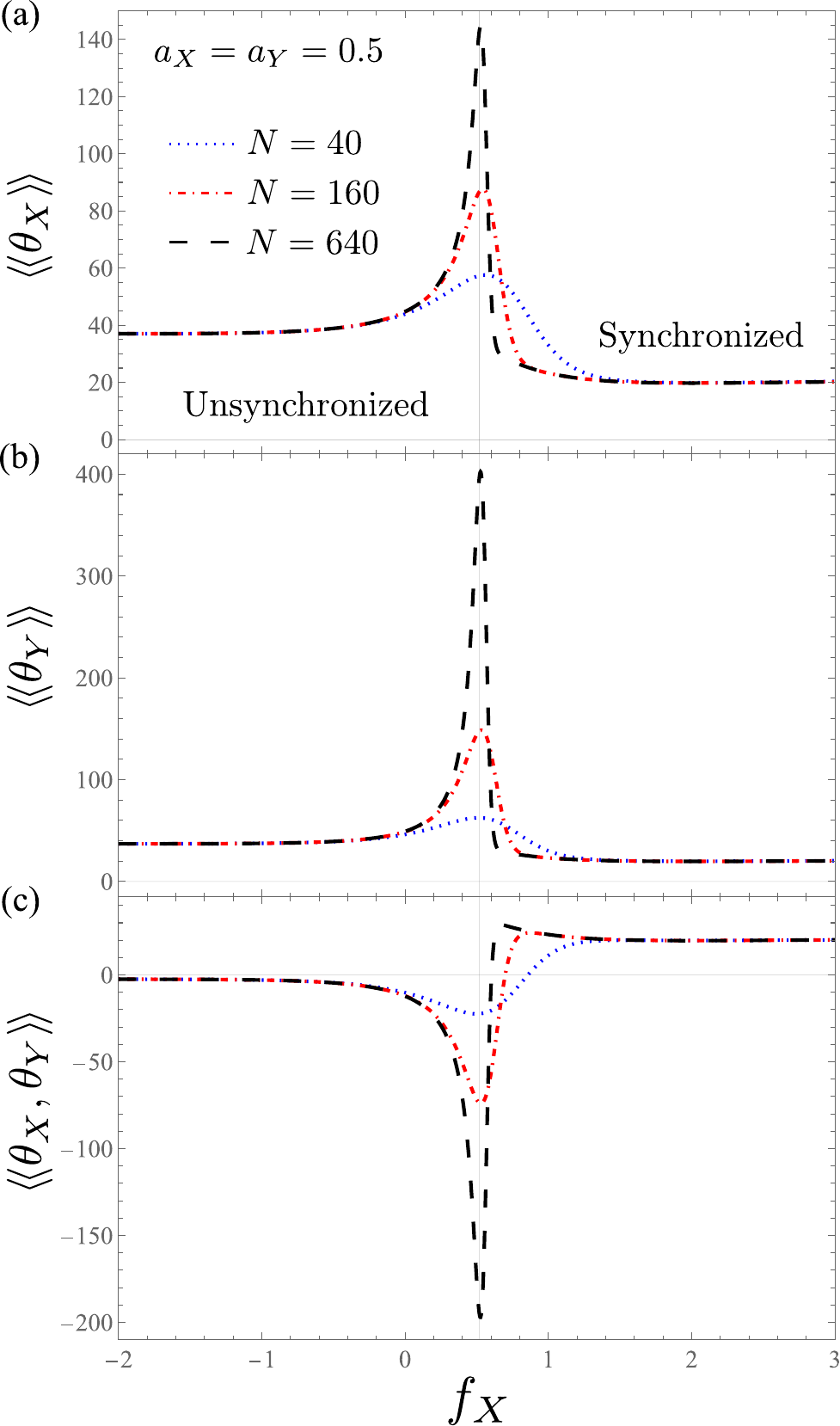}
    \caption{Phase variances $\llangle \theta_\alpha \rrangle$ and the covariance $\llangle \theta_X,\theta_Y \rrangle$ as a function of $f_X$ for the symmetric coupling $a_X=a_Y=0.5$. Note different scales at the $y$ axis. The vertical gray lines denote the phase transition from unsynchronized (left) to synchronized (right) state. Other parameters as in Fig.~\ref{fig:freq}:  $f_Y=2$, $\Gamma_X=\Gamma_Y=1$, $\beta=1$.}
    \label{fig:covariances-symmetric}
\end{figure}
\begin{figure}
    \centering
    \includegraphics[width=0.95\linewidth]{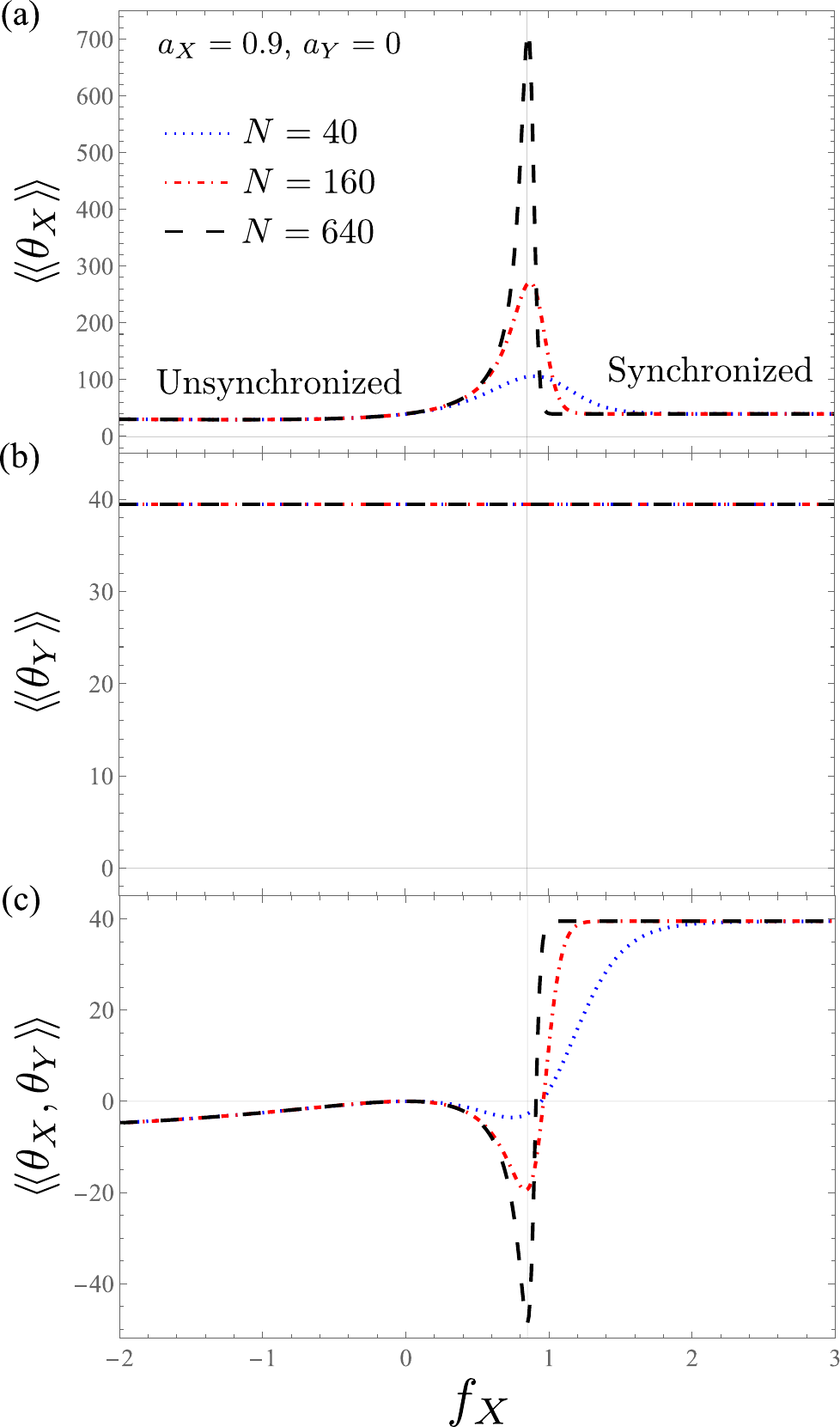}
    \caption{The same plot as Fig.~\ref{fig:covariances-symmetric}, but for asymmetric coupling $a_X=0.9$, $a_Y=0$.}
    \label{fig:covariances-asymmetric}
\end{figure}
We now turn to the behavior of phase covariances. It is plotted in Figs.~\ref{fig:covariances-symmetric} and~\ref{fig:covariances-asymmetric} for the symmetric coupling $a_X=a_Y=0.5$ and the asymmetric coupling $a_X=0.9$, $a_Y=0$, respectively. In the former case, the variances $\llangle \theta_\alpha \rrangle$ increase with the system size $N$ close to the phase transition point, while converge to value independent of $N$ away from the phase transition. This is similar to the behavior of the phase difference variance $\llangle \varphi \rrangle$, and results from the same mechanism. Interestingly, one may also observe that the covariance $\llangle \theta_X,\theta_Y \rrangle$ is negative close to the phase transition point, and tends to $-\infty$ as $N$ grows. To the best of our knowledge, this phenomenon has not previously been reported. Possibly, it can be qualitatively interpreted as follows: Enhancement of $\llangle \varphi \rrangle$ around the phase transition increases the probability that $\Delta \varphi(t) \equiv \Delta \theta_Y(t)-\Delta \theta_X(t)$ is large. It is more probable that such a deviation occurs when $\Delta \theta_Y(t)>0$ and $\Delta \theta_X(t)<0$; otherwise, if $\Delta \theta_X(t)>0$, $\Delta \theta_Y(t)$ would need to be very large, which is less likely. This increases the probability that $\Delta \theta_X(t) \Delta \theta_Y(t)<0$, which by Eq.~\eqref{eq:covariance-def} leads to the negative covariance.

For the asymmetric coupling $a_X=0.9$, $a_Y=0$, we observe a qualitatively similar behavior of $\llangle \theta_X \rrangle$ and $\llangle \theta_X,\theta_Y \rrangle$. However, the variance $\llangle \theta_Y \rrangle$ is now independent of $f_X$, since the dynamics of the $Y$ oscillator is not affected by the state of the $X$ oscillator.

\subsection{Fluctuations of entropy productions}
Finally, to provide some thermodynamic insight, let us consider the fluctuations of entropy production. To that end, we define the stochastic entropy flow to the environment from the subsystem $\alpha \in \{X,Y \}$ as
\begin{align} \label{eq:stochentrprod}
\Sigma_\alpha(t) \equiv \beta n_\alpha(t) f_\alpha \,.
\end{align}
The intensive entropy production rate for the subsystem $\alpha$ equals then
\begin{align}
 \dot{\sigma}_\alpha = \lim_{t \rightarrow \infty} \frac{1}{N t} \left \langle \Sigma_\alpha(t) \right \rangle \,.
\end{align}

The covariance of entropy production in oscillators $\alpha$ and $\alpha'$ is defined as
\begin{align}
   \llangle \dot{\sigma}_\alpha,\dot{\sigma}_{\alpha'} \rrangle \equiv \lim_{t \rightarrow \infty} \frac{1}{Nt} \left \langle \Delta \Sigma_\alpha(t) \Delta \Sigma_{\alpha'}(t) \right \rangle \,,
\end{align}
where $\Delta \Sigma_\alpha(t)=\Sigma_\alpha(t)-\langle \Sigma_\alpha(t) \rangle$. In particular, the variance of entropy production in a single oscillator reads
\begin{align}
\llangle \dot{\sigma}_\alpha \rrangle \equiv \llangle \dot{\sigma}_\alpha,\dot{\sigma}_{\alpha'} \rrangle \,.
\end{align}
The variance of the total entropy production is defined as
\begin{align}
    \llangle \dot{\sigma} \rrangle \equiv \lim_{t \rightarrow \infty} \frac{1}{Nt} \left \langle \left[\Delta \Sigma_X(t)+\Delta \Sigma_Y(t)  \right]^2 \right \rangle \,,
\end{align}
It can be expressed in terms of local variances and covariances as
\begin{align}
 \llangle \dot{\sigma} \rrangle=\llangle \dot{\sigma}_X \rrangle+\llangle \dot{\sigma}_Y \rrangle+2\llangle \dot{\sigma}_X,\dot{\sigma}_Y \rrangle \,.
\end{align}

\begin{figure}
    \centering
    \includegraphics[width=0.95\linewidth]{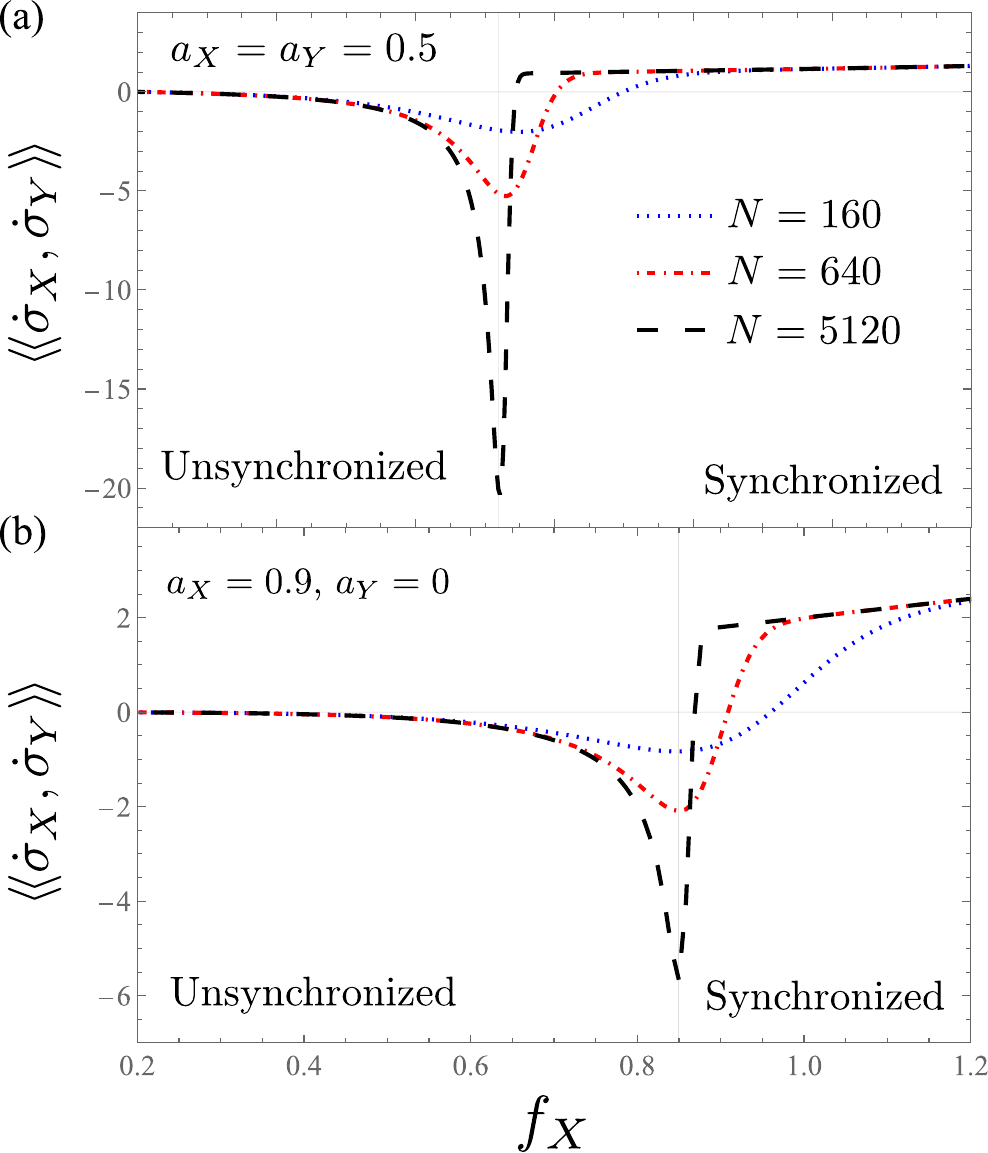}
    \caption{Covariance of local entropy productions $\llangle \dot{\sigma}_X,\dot{\sigma}_Y \rrangle$ as a function of $f_X$ for (a) the symmetric coupling $a_X=a_Y=0.5$ and (b) the asymmetric coupling $a_X=0.9$, $a_Y=0$. Note different scales at the $y$ axis. The vertical gray lines denote the phase transition from unsynchronized (left) to synchronized (right) state. Other parameters are as in Fig.~\ref{fig:freq} ($f_Y=2$, $\Gamma_X=\Gamma_Y=1$, $\beta=1$), but the results are plotted for a smaller range of $f_X$ for better visibility.}
    \label{fig:coventr}
\end{figure}
Importantly, in our model the stochastic entropy production is directly proportional to the stochastic phase difference, $\Sigma_\alpha(t)=N \beta f_\alpha \theta_\alpha(t)/(2 \pi)$ [compare Eqs.~\eqref{eq:stochphase} and~\eqref{eq:stochentrprod}]. Consequently, the entropy production covariances are proportional to phase covariances,
    \begin{align}
    \llangle \dot{\sigma}_\alpha,\dot{\sigma}_{\alpha'} \rrangle &= \frac{\beta^2 \llangle \theta_\alpha,\theta_{\alpha'} \rrangle f_\alpha f_{\alpha'}}{4\pi^2} \,.
    \end{align}
This means that the above-discussed divergent behavior of phase variances near the phase transition point leads to an analogous divergent behavior of variances of local and global entropy productions. In fact, the divergence of entropy production fluctuations close to nonequilibrium phase transitions has been previously reported in Refs.~\cite{nguyen2018phase,oberreiter2021stochastic,remlein2024nonequilibrium}. However, the new observation is that the covariance of local entropy productions, $\llangle \dot{\sigma}_X,\dot{\sigma}_Y \rrangle$, tends to be negative near the synchronization transition and diverges  to $-\infty$ for $N \rightarrow \infty$ (see Fig.~\ref{fig:coventr}). This is because in our model the synchronization transition occurs only when $f_X$ and $f_Y$ are of the same signs (see Secs.~\ref{subsec:nochanges} and~\ref{subsec:nodemons}). Therefore, near the synchronization transition, $\llangle \dot{\sigma}_X,\dot{\sigma}_Y \rrangle$ has the same sign as $\llangle \theta_X,\theta_Y \rrangle$ and follows its behavior presented in Figs.~\ref{fig:covariances-symmetric}~(c) and~\ref{fig:covariances-asymmetric}~(c). Whether this observation can be generalized beyond the model considered to more generic thermodynamically consistent models of synchronization remains an open question.

\section{Mutual information and information flow} \label{sec:inf}

\subsection{Mutual information}

\subsubsection{Definitions}

Finally, we investigate how synchronization influences both the static and dynamic  properties of correlations between the oscillators. To this end, we use information-theoretic measures, which offer a general framework for characterizing correlations that does not depend on the specific physical details of the systems. In particular, to quantify static correlations, we employ mutual information, the standard measure of statistical dependence between two random variables. It is defined as
\begin{align} \label{eq:mutinf}
I_{XY} \equiv \sum_{x,y=0}^{N-1} p_{xy} \ln \frac{p_{xy}}{p_x p_y} \geq 0 \,,
\end{align}
where $p_x \equiv \sum_{y=0}^{N-1} p_{xy}$ and $p_y \equiv \sum_{x=0}^{N-1} p_{xy}$. This quantity has previously been used as a way to quantify the degree of synchronization between oscillators in a manner that is independent of their physical implementation~\cite{zhou2002noise,boccaletti2002synchronization,giorgi2012quantum,manzano2013synchronization,ameri2015mutual,zhu2015synchronization}. In particular, Ref.~\cite{ameri2015mutual} suggested that it can be used as a universal order parameter of synchronization. Refs.~\cite{giorgi2013spontaneous,galve2017quantum} expressed reservations about the generality of that conclusion, pointing out that in some models mutual information is not a distinctive signature of synchronization. We thus ask how mutual information behaves in our model and whether it acts as an order parameter of synchronization. 

To calculate mutual information, we recall that for the model considered, the system's symmetry implies $\forall x,y: \, p_x=p_y=1/N$ and $\forall x,y: p_{xy}=p_i/N$ where $p_i$ are the probabilities of the discrete phase differences (see Sec.~\ref{subsec:reduction}). Consequently, mutual information can be expressed solely in terms of probabilities $p_i$,
\begin{align} \label{eq:mutinf-eff}
I_{XY} =\ln N+\sum_{i=0}^{N-1} p_i \ln p_i \,.
\end{align}
As a result, mutual information is constrained as
\begin{align} \label{eq:mutinf-bounds}
I_{XY} \in [0,\ln N] \,,
\end{align}
with the limit $I_{XY}=0$ obtained for a uniform distribution of phase difference probabilities, $\forall i: p_i=1/N$, and the limit $I_{XY}=\ln N$ obtained for a system occupying a single relative-phase state $i$. This implies that mutual information can scale at most logarithmically with the system size $N$.

\subsubsection{Synchronized state}

We now aim to analytically (or semi-analytically) characterize the behavior of mutual information for large $N$, which will be compared with numerical results for finite $N$. To do this, we replace the probabilities $p_i$ with a continuous probability density $\rho(\varphi)=N p_i/(2\pi)$. Inserting this into Eq.~\eqref{eq:mutinf-eff}, we can approximate mutual information as
\begin{align} \label{eq:mutinfintegral}
I_{XY} \approx \int_0^{2 \pi} \rho(\varphi) \ln [2\pi \rho(\varphi)] d \varphi \,,
\end{align}
which becomes asymptotically exact for $N \rightarrow \infty$. We first focus on the synchronized state, where in the deterministic limit the relative phase $\varphi$ relaxes to a deterministic fixed point $\varphi^*$ given by Eq.~\eqref{eq:fixedpointadler}. Then, employing the Langevin equation~\eqref{eq:langevin} in the limit of $N \rightarrow \infty$, the probability density $\rho(\varphi)$ converges to a Gaussian distribution around that fixed point~\cite{FalascoReview,lax1960fluctuations},
\begin{align} \label{eq:gaussian}
\rho(\varphi) \overset{N \rightarrow \infty}{=} \frac{1}{\sqrt{2 \pi \mathcal{V}/N}} \exp \left[-\frac{(\varphi-\varphi^*)^2 N}{2 \mathcal{V}} \right] \,,
\end{align}
where $\mathcal{V}=\lim_{N \rightarrow \infty} N \langle (\varphi-\varphi^*)^2 \rangle$ is the scaled variance of the distribution. The latter is given by the stationary solution of the Lyapunov equation
\begin{align}
d_t \mathcal{V}=2 \mathcal{V}\partial_\varphi F(\varphi^*)+2D(\varphi^*)=0 \,,
\end{align}
where $F(\varphi)$ and $D(\varphi)$ are the drift and diffusion terms given by Eqs.~\eqref{eq:adler} and~\eqref{eq:diffusion}. The solution yields
\begin{align} \nonumber
&\mathcal{V}=2\pi^2 \frac{\Gamma_X (1+a_X \omega/K)+\Gamma_Y (1-a_Y \omega/K)}{ K \sqrt{1-\omega^2/K^2}} \,.
\end{align}
Inserting this into Eq.~\eqref{eq:mutinfintegral}, we obtain the desired expression for the mutual information,
\begin{figure}
    \centering
    \includegraphics[width=0.9\linewidth]{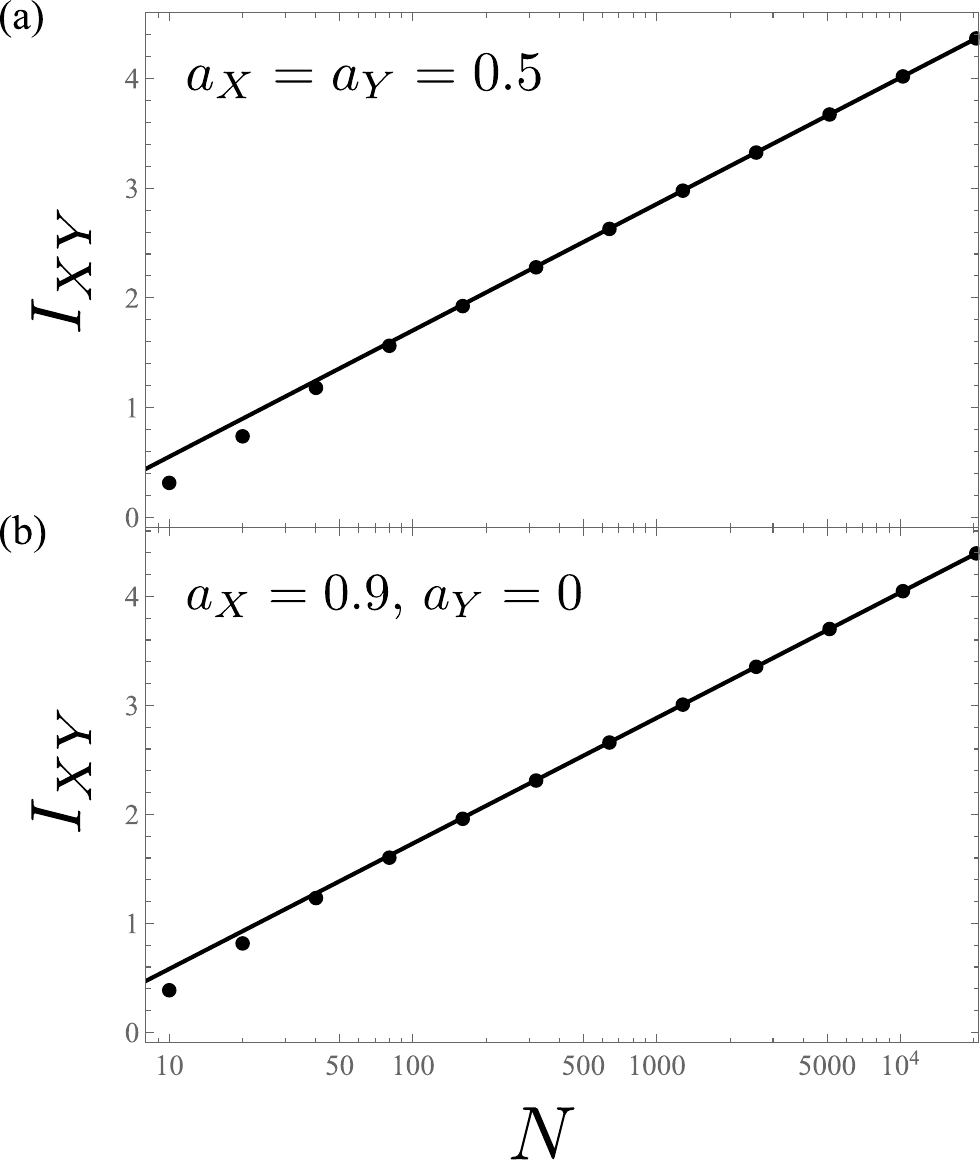}
    \caption{Scaling of mutual information $I_{XY}$ with the system size $N$, plotted in the log-linear scale, for (a) the symmetric coupling $a_X=a_Y=0.5$ and (b) the asymmetric coupling $a_X=0.9$, $a_Y=0$. Dots represent the exact results and the solid line represents the asymptotic expression~\eqref{eq:asympinfsync}. We take $f_X=3$ and other parameters as in Fig.~\ref{fig:freq}:  $f_Y=2$, $\Gamma_X=\Gamma_Y=1$, $\beta=1$.}
    \label{fig:scaling-information-synchronized}
\end{figure}
\begin{align} \label{eq:asympinfsync}
I_{XY} \overset{N \rightarrow \infty}{=}  \frac{\ln N}{2}+\frac{1}{2} \ln\frac{2 \pi}{e \mathcal{V}} \,.
\end{align}
This implies that mutual information scales logarithmically with $N$ in the synchronized state. The validity of this expression is demonstrated in Fig.~\ref{fig:scaling-information-synchronized}. As shown, the approximation works very well already for $N \approx 100$ (a) or $N=50$ (b), and becomes asymptotically exact for large $N$. We note in passing that similar logarithmic scaling of mutual information between two subsystems have been observed at certain equilibrium phase transitions~\cite{wilms2012finite}.

\subsubsection{Unsynchronized state}
\begin{figure}
    \centering
    \includegraphics[width=0.95\linewidth]{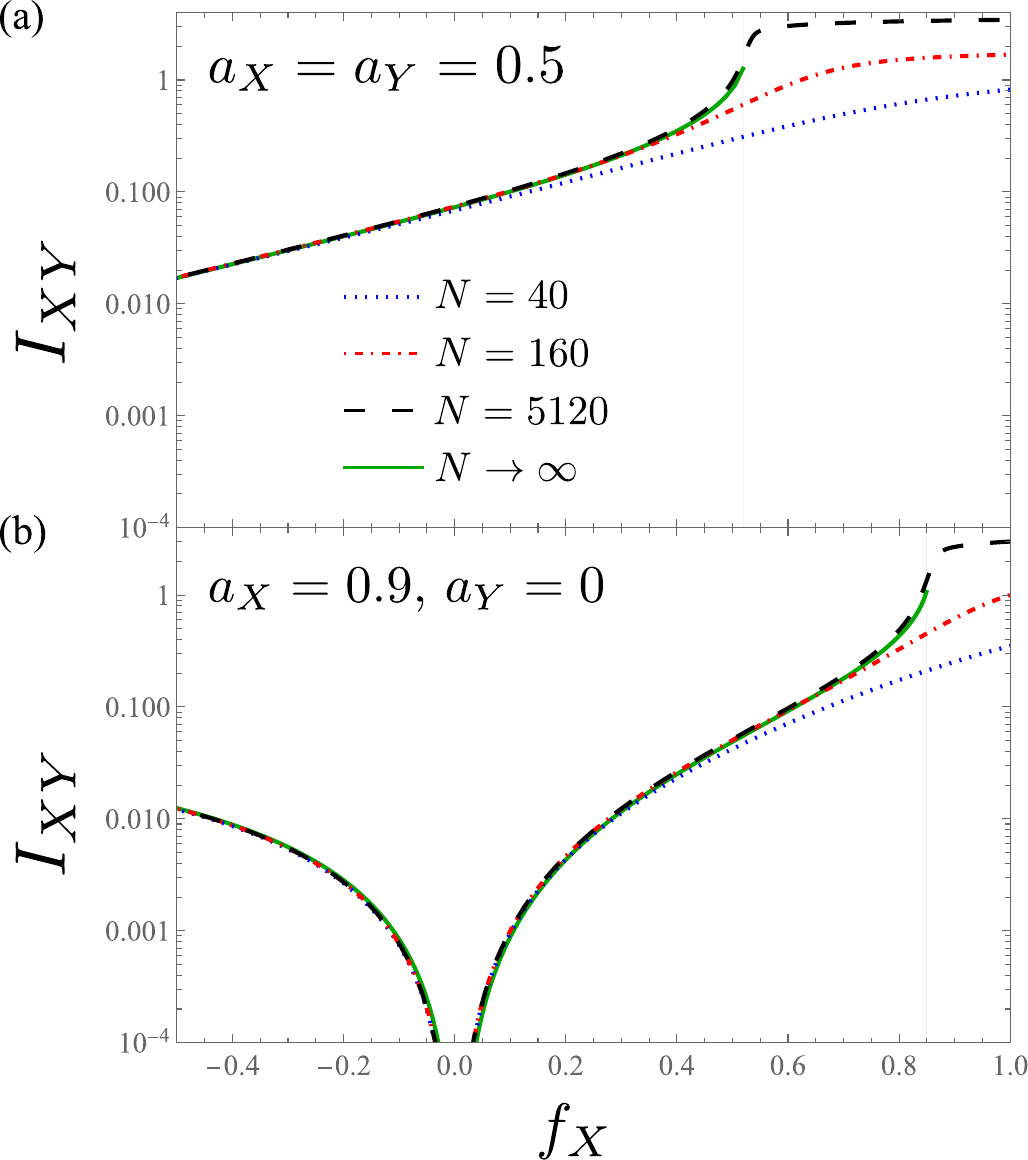}
    \caption{Mutual information $I_{XY}$ as a function of $f_X$, plotted in the logarithmic scale, for (a) the symmetric coupling $a_X=a_Y=0.5$ and (b) the asymmetric coupling $a_X=0.9$, $a_Y=0$. The vertical gray lines denote the phase transition from unsynchronized (left) to synchronized (right) state. Results for finite $N$ are obtained using master equation approach, while results for $N \rightarrow \infty$ are given by Eq.~\eqref{eq:mutinfintegral} with $\rho(\varphi)$ given by Eq.~\eqref{eq:probden-unsync}. In (b), $I_{XY}=0$ for $f_X=0$. Other parameters as in Fig.~\ref{fig:freq}:  $f_Y=2$, $\Gamma_X=\Gamma_Y=1$, $\beta=1$.}
    \label{fig:information-unsync}
\end{figure}
We now turn to the unsynchronized state. In this case, we use the fact that in the stationary state the probability current $\rho(\varphi) F(\varphi)$ has to be equal for all $\varphi$. Consequently, the probability density scales asymptotically as~\cite{dykman1993stationary,vance1996fluctuations,ge2012landscapes}
\begin{align} \label{eq:probden-unsync}
\rho(\varphi) \overset{N \rightarrow \infty}{=} \frac{1}{T F(\varphi)} \,,
\end{align}
where $T$ is the period of relative phase evolution given by Eq.~\eqref{eq:periodphasedif}, which provides the normalization of the probability density. The mutual information can then be numerically evaluated using Eq.~\eqref{eq:mutinfintegral}. Crucially, since the probability density $\rho(\varphi)$ is of the order $O(1)$, mutual information in the unsynchronized state becomes asymptotically an intensive quantity, independent of system size $N$. This is illustrated in Fig.~\ref{fig:information-unsync}. As shown there, far enough from transition to synchronization, mutual information is already approximately size-independent for relatively small $N \gtrapprox 40$ and agrees with the predictions of our theory. In contrast, close to transition to synchronization, the agreement is observed only for large $N$. We also observe that for the asymmetric coupling $a_X=0.9$, $a_Y=0$, mutual information is equal to $0$ for $f_X=0$. This may be explained as follows: First, the dynamics (and thus the state) of $Y$ oscillator does not depend on the state of $X$ oscillator. Second, the state of the $Y$ oscillator affects only the kinetics of the $X$ oscillator (i.e., the symmetric part of transition rates\footnote{See the paragraph preceding Sec.~\ref{subsec:reduction}.}), which does not affect the state of $X$ oscillator at thermal equilibrium $f_X=0$. Consequently, the states of both oscillators are mutually independent, so that $I_{XY}=0$.

\subsubsection{Crossover from logarithmic to intensive scaling}
\begin{figure}
    \centering
    \includegraphics[width=0.95\linewidth]{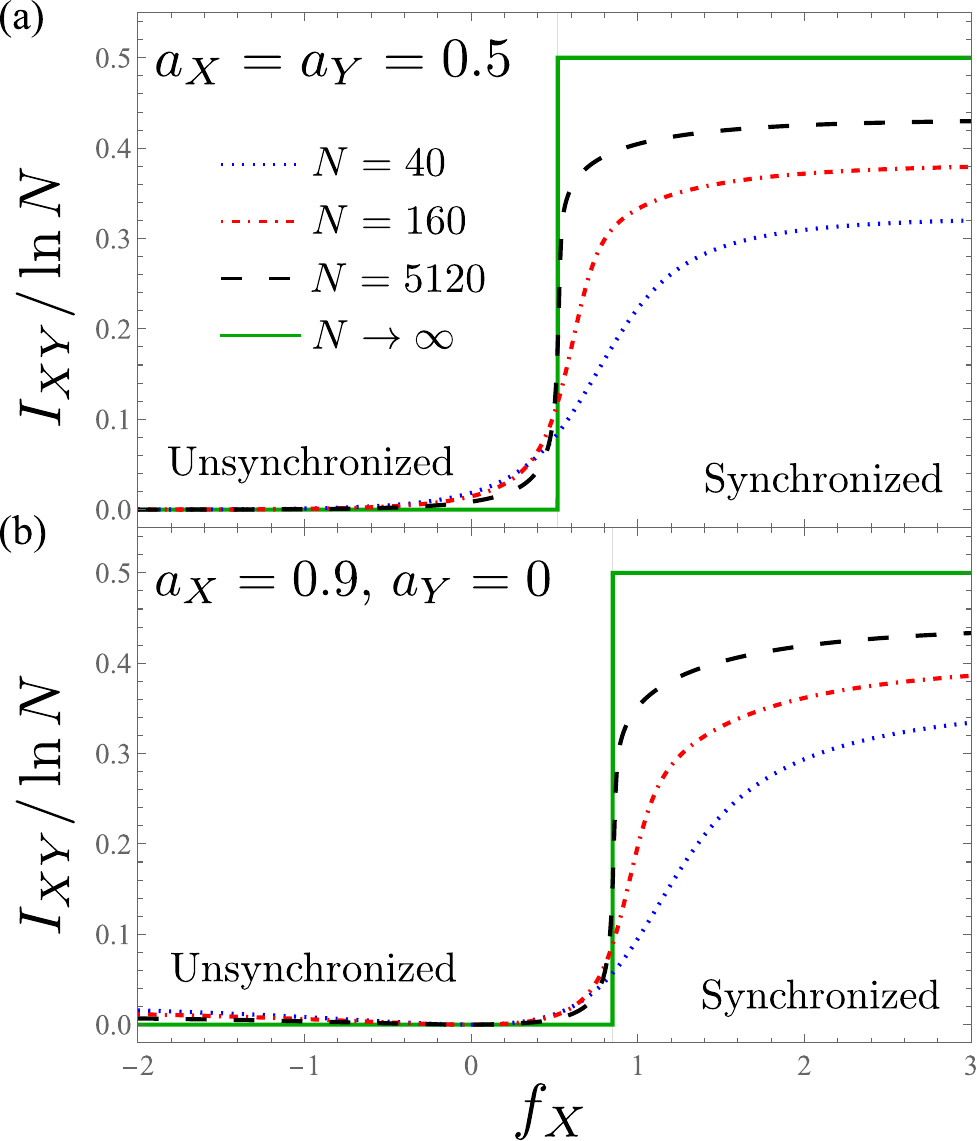}
    \caption{$I_{XY}/\ln N$ as a function of $f_X$ for (a) the symmetric coupling $a_X=a_Y=0.5$ and (b) the asymmetric coupling $a_X=0.9$, $a_Y=0$. Sharp jump for $N \rightarrow \infty$ indicates a transition from the unsynchronized (left) to the synchronized (right) state. Other parameters as in Fig.~\ref{fig:freq}:  $f_Y=2$, $\Gamma_X=\Gamma_Y=1$, $\beta=1$.}
    \label{fig:information}
\end{figure}
Our results imply that mutual information exhibits two distinct scaling behaviors: it scales logarithmically with $N$ in the synchronized state, while it becomes intensive (independent of $N$) in the unsynchronized state. To illustrate that crossover, in Fig.~\ref{fig:information} we plot $I_{XY}/(\ln N)$ as a function of $f_X$ across the nonequilibrium phase transition from unsynchronized to synchronized state. As shown, in the unsynchronized state, this quantity gradually vanishes with $N$ as $\propto 1/\ln N$. In contrast, in the synchronized state it converges with $N$ to a finite value $1/2$ [see Eq.~\eqref{eq:asympinfsync}]. We note that this convergence is quite slow, due to the slow decay of the second term on the right-hand side of Eq.~\eqref{eq:asympinfsync} divided by $\ln N$. This  shows that for $N \rightarrow \infty$ the quantity $I_{XY}/\ln N$ acts as a genuine order parameter of synchronization, behaving discontinuously at the phase transition point. 

\subsection{Information flow} \label{subsec:infflow}
\subsubsection{Definitions}
We now turn to the dynamical properties of correlations. To characterize them, we employ
the information flow from the oscillator $Y$ to $X$~\cite{horowitz2014thermodynamics}
\begin{align} \label{eq:infflow-def}
\mathcal{I} \equiv \sum_{x,y=0}^{N-1} \left(W^y_{x+1,x} p_{xy} -W^y_{x,x+1} p_{x+1,y}\right)\ln \frac{p_{x+1,y}}{p_{xy}} \,.
\end{align}
This quantity characterizes the mutual dependence of the dynamics of two oscillators: it is positive when the dynamics of the oscillator $X$ is more strongly affected by the state of oscillator $Y$ than vice versa. It attracted a peculiar attention in the context of autonomous Maxwell demons discussed in Sec.~\ref{subsec:nodemons}, i.e., the bipartite stochastic systems in which one of the subsystems can continuously reduce the entropy of its environment ($\dot{\sigma}_\alpha<0$) due to autonomous control by the other subsystem. In that context, the information flow constrains the demon operation via the generalized second law of thermodynamics $N \dot{\sigma}_X-\mathcal{I} \geq 0$, $N \dot{\sigma}_Y+\mathcal{I} \geq 0$~\cite{horowitz2014thermodynamics}.\footnote{We recall that in our paper $\dot{\sigma}_\alpha$ denotes the intensive entropy production rate, i.e., divided by system size $N$.} More generally, this quantity has been used to characterize the cooperative behavior of multicomponent molecular machines, where free energy transduction includes both energetic and informational components~\cite{lathouwers2022internal,leighton2023inferring,grelier2024unlocking,leighton2024information,ehrich2023energy,leighton2025flow}.

\subsubsection{Synchronized state}
As previously, our goal now is to describe the asymptotic behavior of the information flow for large $N$. To that end, we employ the previously noted system's symmetry $\forall x,y: \, p_x=p_y=1/N$ and $\forall x,y: p_{xy}=p_i/N$ to rewrite Eq.~\eqref{eq:infflow-def} as
\begin{align} \label{eq:infflow-red}
\mathcal{I} =\sum_{i=0}^{N-1} (W^{X}_{i-1,i} p_i -W^{X}_{i,i-i} p_{i-1})\ln \frac{p_{i-1}}{p_{i}} \,.
\end{align}
We now focus on the synchronized state. In that context, it is convenient to rewrite Eq.~\eqref{eq:infflow-red} as
\begin{align} \nonumber
\mathcal{I}=
  &-\sum_{i=0}^{N-1} \left(W^{X}_{i-1,i} p_i -W^{X}_{i,i-i} p_{i-1}\right) \ln p_i \\ \nonumber
  &+\sum_{i=0}^{N-1} \left(W^{X}_{i-1,i} p_i -W^{X}_{i,i-i} p_{i-1}\right) \ln p_{i-1} \\ \nonumber
= &-\sum_{i=0}^{N-1} \left(W^{X}_{i-1,i} p_i -W^{X}_{i,i-i} p_{i-1}\right) \ln p_i\\ \nonumber
  &+\sum_{i=0}^{N-1} \left(W^{X}_{i,i+1} p_{i+1} -W^{X}_{i+i,i} p_{i}\right) \ln p_{i}\\
= & \sum_{i=0}^{N-1} \sum_{\pm} \left( W^X_{i,i\pm 1} p_{i \pm 1}-W^X_{i \pm 1,i} p_i \right) \ln p_i \,,
\end{align}
where in the second step we replaced the index $i \rightarrow i+1$ in the second sum. Using Eq.~\eqref{eq:masteq-eff}, we can identify the expression $\sum_{\pm} ( W^X_{i,i\pm 1} p_{i \pm 1}-W^X_{i \pm 1,i} p_i)$ in the last term of the equation above as the rate of change of the probability $p_i$ induced by the jumps in the oscillator $X$. Denoting this term as $\dot{p}_i^X$, we obtain
\begin{align} \nonumber
\mathcal{I}=\sum_{i=0}^{N-1} \dot{p}_i^X \ln p_i \,.
\end{align}
In the continuous limit, this yields an analogous expression \begin{align} \label{eq:infflow-sync-int}
\mathcal{I}=\int_0^{2 \pi} d\varphi \dot{\rho}^X(\varphi) \ln [2 \pi \rho(\varphi)] \,,
\end{align}
where, correspondingly, $\dot{\rho}^X(\varphi)$ is the rate of change of probability density due to transitions in the oscillator $X$. Using the Kramers-Moyal expansion of the master equation~\cite{kramers1940brownian,moyal1949stochastic}, it can be expressed as
\begin{align} \label{eq:probden-ev-x} \nonumber
\dot{\rho}^X(\varphi)=&-\frac{\partial}{\partial \varphi} \left[F_X(\varphi) \rho(\varphi) \right]+\frac{1}{N} \frac{\partial^2}{\partial \varphi^2} \left[D_X(\varphi) \rho(\varphi) \right] \\
&+O(N^{-2}) \,,
\end{align}
where
\begin{subequations}
\begin{align}
F_X(\varphi) &\equiv 2\pi \left[w_+^{X}(\varphi)-w_-^{X}(\varphi) \right]=-\Omega_X-K_X \sin(\varphi) \,, \\ \nonumber
D_X(\varphi)  &\equiv \frac{4\pi^2}{2} \left[w_+^{X}(\varphi)+w_-^{X}(\varphi) \right] \\ &=2 \pi^2 \Gamma_X \left[1+a_X \sin(\varphi) \right] \,,
\end{align}
\end{subequations}
are the contributions to the drift and diffusion term related to transitions in the subsystem $X$, and $w^\alpha_\pm(\varphi)$ are the intensive transition rates given by Eq.~\eqref{eq:intrates}. Following Ref.~\cite{freitas2023information}, here we assume that the information flow -- analogously to mutual information -- is determined by weak Gaussian fluctuations around the deterministic fixed point $\varphi^*$. Therefore, we truncate the Kramers-Moyal expansion after the diffusion term. Assuming the Gaussian probability density around the fixed point given by Eq.~\eqref{eq:gaussian}, we can express the derivatives of the probability density appearing in Eq.~\eqref{eq:probden-ev-x} as
\begin{subequations}
\begin{align}
\frac{\partial}{\partial \varphi} \rho(\varphi) & =-N(\varphi-\varphi^*)\rho(\varphi)/\mathcal{V} \,, \\
\frac{\partial^2}{\partial \varphi^2} \rho(\varphi) &=-N\rho(\varphi)/\mathcal{V}+N^2(\varphi-\varphi^*)^2\rho(\varphi)/\mathcal{V}^2 \,.
\end{align}
\end{subequations}
Since the integral~\eqref{eq:infflow-sync-int} is dominated by the behavior around $\varphi^*$, we further employ the Taylor expansion of the drift term,
\begin{align}
F_X(\varphi)=&F_X(\varphi^*)+ (\varphi-\varphi^*) \partial_\varphi F_X(\varphi^*) +O[(\varphi-\varphi^*)^2] \,.
\end{align}
Inserting the above expressions into Eq.~\eqref{eq:probden-ev-x} and evaluating the integral~\eqref{eq:infflow-sync-int}, we obtain
\begin{align} 
\mathcal{I} &=-\partial_\varphi F(\varphi^*)-D_X (\varphi^*)/\mathcal{V}+O(1/N) \,.
\end{align}
This yields an asymptotic expression for the information flow,
\begin{align} \label{eq:infflowsync} 
\lim_{N \rightarrow \infty} \mathcal{I} =\frac{\Omega_X K_X/\Gamma_X-\Omega_Y K_Y/\Gamma_Y}{\Omega_X/\Gamma_X+\Omega_Y/\Gamma_Y} \cos(\varphi^*) \,.
\end{align}
Importantly, the above equation implies that information flow is intensive in the synchronized state, i.e., it saturates at a finite value independent of $N$. We note that similar intensive scaling has been previously observed for electronic~\cite{freitas2023information} or chemical~\cite{bilancioni2023chemical} models of Maxwell demons.

\subsubsection{Unsynchronized state}
We now turn to the unsynchronized state. In that case, we apply the following continuous limit to Eq.~\eqref{eq:infflow-red}
\begin{subequations}
    \begin{align}
&W^{X}_{i-1,i} p_i -W^{X}_{i,i-i} p_{i-1} \rightarrow -F_X(\varphi) \rho(\varphi)+O(1/N) \,, \\ \nonumber
&\ln \frac{p_{i-1}}{p_i} \rightarrow \ln \frac{\rho(\varphi-2 \pi/N)}{\rho(\varphi)}\\ &=-2\pi \partial_\varphi \rho(\varphi)/[N \rho(\varphi)]+O(N^{-3}) \,.
\end{align}
\end{subequations}
As a result, we obtain
\begin{align}
\mathcal{I}=2 \pi \int_0^{2 \pi} F_X(\varphi) \partial_\varphi \rho(\varphi) d \varphi+O(1/N) \,,
\end{align}
with $\rho(\varphi)$ given by Eq.~\eqref{eq:probden-unsync}. By direct calculation, we verified that the above integral yields 0. Qualitatively, this may be related to the fact that information flow is related to rectification of fluctuations, and thus should not be determined by the integral above, which is expressed solely in terms of deterministic dynamics. As a result, the information flow is of order $O(1/N)$ and thus asymptotically vanishes,
\begin{align} \label{eq:infflowunsync}
\lim_{N \rightarrow \infty} \mathcal{I}=0 \,.
\end{align}
Consequently, since for $N \rightarrow \infty$ the information flow is finite in the synchronized state and vanishes in the unsynchronized state, it acts as an order parameter of synchronization (similarly to the quantity $I_{XY}/\ln N$).

\subsubsection{Numerical results}
\begin{figure}
    \centering
    \includegraphics[width=0.95\linewidth]{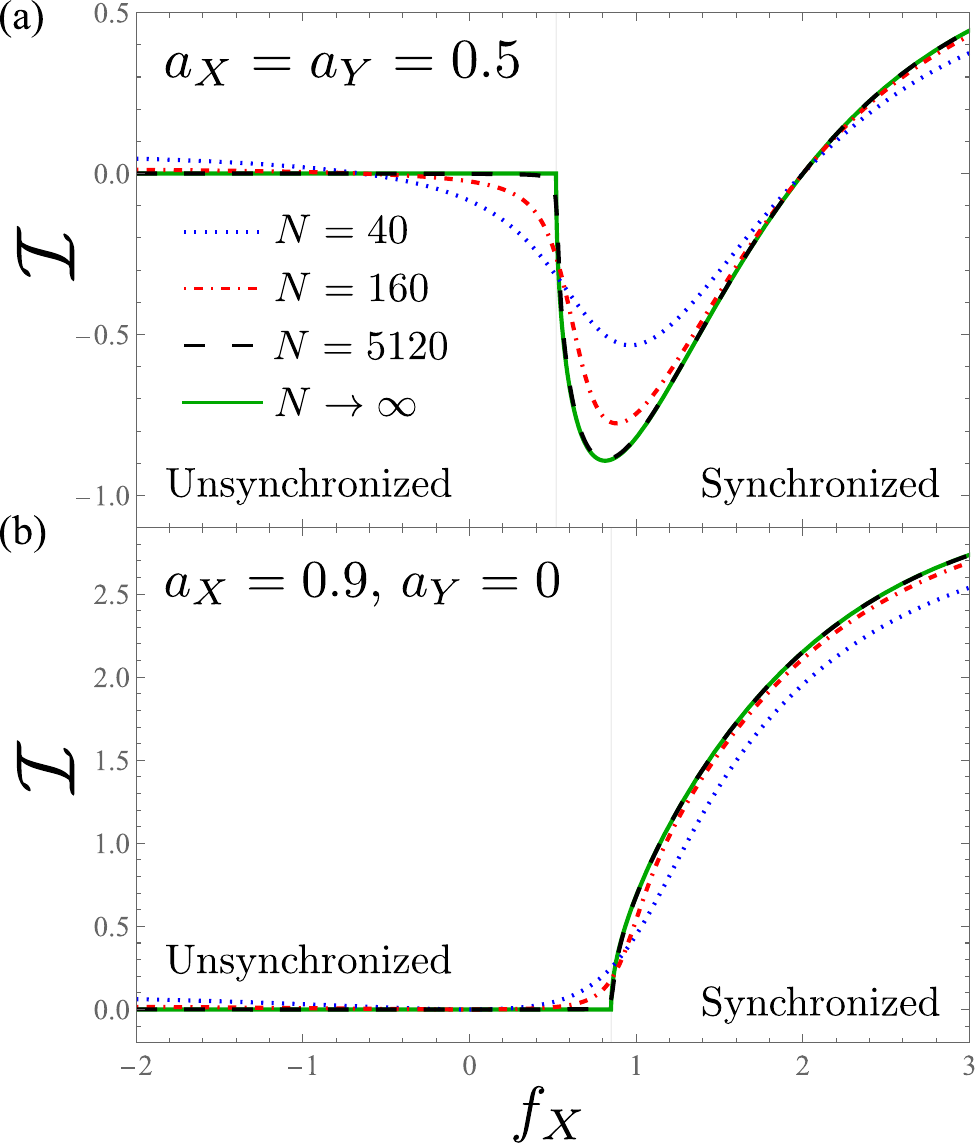}
    \caption{Information flow $\mathcal{I}$ as a function of $f_X$ for (a) the symmetric coupling $a_X=a_Y=0.5$ and (b) the asymmetric coupling $a_X=0.9$, $a_Y=0$. Note different scales on the $y$-axis in (a) and (b). The vertical gray lines denote the phase transition from unsynchronized(left) to synchronized (right) state. Results for finite $N$ are obtained using master equation approach, while results for $N \rightarrow \infty$ are given by the asymptotic expressions~\eqref{eq:infflowsync} and~\eqref{eq:infflowunsync}. Other parameters as in Fig.~\ref{fig:freq}:  $f_Y=2$, $\Gamma_X=\Gamma_Y=1$, $\beta=1$.}
    \label{fig:infflow}
\end{figure}
To verify our theory, we compare the asymptotic expressions~\eqref{eq:infflowsync} and~\eqref{eq:infflowunsync} with the numerical results for finite system sizes. As shown in Fig.~\ref{fig:infflow}, the information flow gradually converges with $N$ to the predictions of our theory. For large $N=5120$, the agreement is nearly perfect. We also note that for the symmetric coupling $a_X=a_Y$ the information flow changes sign for $f_X=f_Y$. In this point, the oscillators are identical and influence each other in the same way, so the information flow vanishes. Away from this point, in the synchronized state, the information tends to flow from the less-dissipating oscillator (i.e., the one with smaller force). This is similar to situation encountered in autonomous Maxwell demons, where the information flows from demon (subsystem with negative dissipation) to feedback device which dissipates the energy, although our setup does not work as a Maxwell demon. In contrast, for the asymmetric case with $a_Y$, the information always flows from $Y$ to the $X$ oscillator ($\mathcal{I}>0$), because the dynamics of the $Y$ oscillator is not affected by the state of the $X$ oscillator.

\subsubsection{Generality of observations} \label{subsec:generality-infflow}

Finally, let us comment on whether we expect our conclusion that information flow can serve as a witness of synchronization to be general or merely a specific feature of the model considered here.  Recall that, in our model, the crossover in the asymptotic information flow $\lim_{N \rightarrow \infty} \mathcal{I}$---from zero in the unsynchronized state to a finite, intensive value in the synchronized state---results from a qualitative change in the properties of the probability density of the phase difference $\rho(\varphi)$. In the unsynchronized state, this distribution is spread over the entire interval $[0,2 \pi]$ and is governed by deterministic dynamics~\eqref{eq:adler}, resulting in vanishing of the information flow. By contrast, in the synchronized phase it becomes a sharply peaked Gaussian distribution determined by weak fluctuations around a fixed point. In this case, $\lim_{N \rightarrow \infty} \mathcal{I}$ is finite, indicating that correlations of fluctuations are necessary for the presence of information flow.

We therefore expect that our conclusion can be generalized to other systems of coupled one-dimensional phase oscillators governed by equations of motion for the phase difference similar to Eq.~\eqref{eq:langevin}, including both continuous-phase noisy variants of the Kuramoto model~\cite{acebron2005kuramoto} and discrete-phase oscillators considered in Refs.~\cite{escaff2016synchronization,jorg2017stochastic,zhang2020energy,zhang2025altruistic}. However, in general, we do not expect the information flow to be a distinctive or universal witness of synchronization. In fact, it can also be finite and intensive in the macroscopic limit of systems that do not exhibit oscillations at all but instead relax to fixed points, such as electronic systems~\cite{freitas2023information} or chemical reaction networks~\cite{bilancioni2023chemical}. In such models---similarly to the synchronized state in our system---the asymptotic value of the information flow is determined by weak Gaussian fluctuations around the fixed point.

\section{Conclusions} \label{sec:concl}
Our study has revealed a rich dynamic and thermodynamic behavior of the model considered. At the dynamical level, we have shown that in the thermodynamic limit $N \rightarrow \infty$ the observed frequencies of the oscillators behave continuously but nonanalytically at the synchronization transition, showing it to be a continuous phase transition. In particular, the detuning of the observed frequencies obeys a universal critical behavior with a critical exponent $1/2$. For finite system sizes, we demonstrated a universal scaling behavior of the frequency detuning between oscillators close to synchronization transition. In particular, the response of frequency detuning to system parameters has been shown to obey a polynomial scaling $\propto N^{1/3}$ close to synchronization transition.

Analyzing the system thermodynamics, we made three main observations: (1) In the deterministic description, the response of the system to forces $f_\alpha$ is intrinsically nonlinear. The linear-response regime is well-defined only at the level of stochastic description, and its range of applicability shrinks as $1/N$. (2) The synchronization transition is not governed by any extremum dissipation principle: synchronization may either enhance or reduce dissipation, depending on system parameters. (3) In the deterministic limit $N \rightarrow \infty$, the system cannot operate as an autonomous Maxwell demon, as previously shown for the models considered in Refs.~\cite{freitas2022maxwell,freitas2023information,bilancioni2023chemical}.

We further investigated the properties of the variances and covariances of the phases and the entropy production. We have shown that the variances tend to diverge with system size $N$ close to the synchronization transition, a behavior typical for continuous phase transitions~\cite{nguyen2018phase,oberreiter2021stochastic,kewming2022diverging,remlein2024nonequilibrium,ptaszynski2024critical}. In particular, phase difference variance has been shown to obey a universal scaling behavior, growing as $N^{2/3}$. Interestingly, close to the synchronization transition, the covariances of phases and local entropy productions go to $-\infty$ as $N \rightarrow \infty$, a phenomenon not previously reported. 

Finally, we analyzed the scaling behavior of mutual information and the information flow between the oscillators. We have shown that mutual information undergoes a crossover from logarithmic scaling with $N$ in the synchronized state to intensive ($N$-independent) scaling in the unsynchronized state. The information flow is intensive but finite in the synchronized state, while tends to $0$ with $N$ in the unsynchronized state. This shows that, in our model, both quantities act as order parameters of the synchronization transition (as suggested in Ref.~\cite{ameri2015mutual}).

This raises the question which part of our conclusions can be generalized beyond the toy model considered to systems of coupled limit-cycle oscillators (e.g., chemical oscillators~\cite{nandi2007effective,nandi2010intrinsic}). First, by means of counterexample, our work illustrates that there is no universal (model-independent) extremum dissipation principle governing the synchronization phase transition. We note that certain synchronization models may still exhibit some model-specific principles, where, regardless of the parameters, synchronization always enhances~\cite{zhang2020energy,guislain2023nonequilibrium} or reduces~\cite{izumida2016energetics,
herpich2018collective,meibohm2024minimum,meibohm2024small,gopal2025dissipation} dissipation. However, in other models (such as our model or the even-coupling scenario in Ref.~\cite{izumida2016energetics}) synchronization may either enhance or reduce dissipation, depending on the parameters. Second, our conclusion about the scaling of mutual information is probably applicable to systems of limit-cycle oscillators. This is because, in such systems, the probability density of a single oscillator is concentrated along the limit-cycle trajectory, making the oscillators effectively one-dimensional, similarly to our model. After applying phase reduction, the dynamics of their phase difference is described using the Langevin equation similar to Eq.~\eqref{eq:langevin}, which provided the basis for our analytic results~\cite{amro2015phase}. For the same reason, conclusions about the scaling of responses or fluctuations of phase difference, or negative phase covariance close to the synchronization transition, are possibly also relevant for systems of coupled limit-cycle oscillators. We note that in the latter context the phase difference can be defined using the stochastic phase definitions from Ref.~\cite{freund2003frequency}. It is less obvious whether the observation of negative covariances of local entropy productions close to synchronization transition can be generalized to coupled limit-cycle oscillators. In such systems, stochastic entropy production is no longer directly proportional to the stochastic phase difference, and the role of energy transduction between the oscillators or fluctuations transverse to the limit cycle may be important. Also, as elaborated in Sec.~\ref{subsec:generality-infflow}, the conclusion that information flow acts as an order parameter of synchronization is probably not general, as this flow can also be finite in the macroscopic limit of models which do not exhibit oscillations~\cite{freitas2023information,bilancioni2023chemical}.

\acknowledgments

M.C.\ and K.P.\ acknowledge the financial support of the National Science Centre, Poland, under project No.\ 2023/51/D/ST3/01203, and M.E.\ the Fonds National de la Recherche-FNR, Luxembourg, under project No.\ C24/MS/18933049/NEQPHASETRANS.  

\section*{Data availability}
The data and source code that support the findings of this article are openly
available at~\cite{zenodo,github}.

\bibliography{bibliography}	
	
\end{document}